\documentclass[journal,10pt,twocolumn,final]{IEEEtran}

\usepackage[T1]{fontenc}
\usepackage[utf8]{inputenc}
\usepackage[nolist]{acronym}
\usepackage{amsmath,amssymb}
\usepackage{mathtools}

\usepackage[linesnumbered,ruled,vlined]{algorithm2e}
\usepackage{algorithmic}
\usepackage{cite}
\usepackage{graphicx}
\usepackage{booktabs}
\usepackage{array,tabularx}
\usepackage{xcolor}
\usepackage[caption=false,font=footnotesize]{subfig}
\graphicspath{{Figures/}}
\usepackage{url}

\begin{document}

\title{Intelligent Carbon Emission-aware Radio Unit Activation Control in Dense Heterogeneous Energy 6G Networks}
\author{\IEEEauthorblockN{Xuanyu Liang\IEEEauthorrefmark{1}, Mostafa Rahmani Ghourtani\IEEEauthorrefmark{1}, Swarna Chetty\IEEEauthorrefmark{1}, Mehdi Sookhak\IEEEauthorrefmark{2}
and Hamed Ahmadi\IEEEauthorrefmark{1}}
\\
\IEEEauthorrefmark{1}School of Physics Engineering and Technology, University of York, United Kingdom\\
\IEEEauthorrefmark{2}Department of Computing Sciences, Texas A\&M University-Corpus Christi, United States of America
}

\markboth{IEEE Transactions on Green Communications and Networking}%
{Liang \MakeLowercase{\textit{et al.}}: Carbon Emission-Aware RU Activation Control}

\maketitle

\begin{abstract}
Base stations account for a dominant share of energy consumption and operational carbon emissions in radio access networks. In \ac{O-RAN} deployments with heterogeneous energy sources, however, minimizing energy consumption alone does not necessarily minimize carbon emissions. In this work, we investigate carbon-aware \ac{DRL} for \ac{RU} sleep control. Under the adopted source-matched operational accounting model, renewable-matched \acp{RU} are assigned zero direct operational emissions, whereas non-renewable-supplied \acp{RU} incur operational emissions. The problem is formulated as a Markov decision process and evaluated using two carbon-aware reward formulations and a conventional energy--QoS reward baseline. We implement and compare \ac{TD3}, \ac{PPO}, and a heuristic method under realistic mobility and 3GPP channel models across multiple renewable-matched RU layouts and traffic loads. Simulation results show that TD3 learns more energy-efficient activation policies than PPO, achieving approximately 3\%--6\% lower energy consumption in representative scenarios. TD3 also achieves 8\%--12\% lower operational emissions than PPO in scenarios with a low proportion of renewable-matched \acp{RU} and up to a 20\% reduction relative to the heuristic baseline under moderate renewable availability and high traffic load. Moreover, the joint energy,carbon and QoS reward achieves approximately 12\%--28\% lower operational emissions than the energy and QoS reward, while maintaining comparable energy consumption.
\end{abstract}


\section{Introduction}

The explosive growth of mobile data traffic has driven the large-scale densification of cellular infrastructures. While dense deployments significantly enhance network capacity and coverage, they simultaneously introduce substantial operational energy demands. Recent reports indicate that \acp{BS} account for approximately 73\%--77\% of the total energy consumption in mobile networks~\cite{thompson2022editorial}, making them the dominant contributors to both operational expenditure and greenhouse gas emissions. A typical 5G macro \ac{BS} consumes between $3.3$ and $9$~kW, with even higher values observed in massive MIMO and mmWave deployments \cite{liang2024energy}. As networks evolve toward ultra-dense 6G architectures, energy efficiency alone is no longer sufficient; reducing carbon emissions associated with network operation has become a central sustainability objective.

Traditional optimization efforts have primarily focused on minimizing instantaneous power consumption. However, energy efficiency does not necessarily translate into carbon efficiency \cite{ahmadi2025towards}. With the increasing adoption of renewable energy sources at cellular sites, modern radio access networks are progressively transitioning toward hybrid energy supply models, where some infrastructure elements are powered by renewable sources (e.g., solar or wind), while others remain dependent on carbon-intensive grid electricity. In such heterogeneous energy environments, minimizing total energy usage may not lead to minimal carbon emissions. Instead, sustainable network operation requires explicit consideration of the carbon intensity of different energy sources. This shifts the optimization paradigm from purely energy-aware control toward carbon-aware and sustainability-driven orchestration.

\ac{O-RAN} provides a promising architectural foundation for enabling such sustainability-oriented control. By disaggregating the traditional monolithic \ac{BS} into \ac{CU}, \ac{DU}, and \ac{RU} components, O-RAN introduces programmability and openness through standardized interfaces~\cite{O-RanArchitecture}. Among these components, \acp{RU} are responsible for lower-PHY processing and RF transmission, and thus constitute a significant portion of radio access energy consumption~\cite{abubakar2023energy}. Since \acp{RU} are typically dimensioned for peak traffic, placing a subset into sleep mode during lower demand periods provides operational flexibility. In hybrid renewable--grid deployments, such flexibility enables differentiated activation strategies, where renewable-powered RUs can be prioritized over grid-powered ones, facilitating structural carbon reduction beyond conventional energy-saving mechanisms.

The introduction of the \ac{RIC} further enhances O-RAN's capability for sustainability-driven control. The \ac{Non-RT RIC} provides long-term policy guidance and optimization objectives via the A1 interface \cite{O-RanA1Interface}, while the \ac{Near-RT RIC} executes fine-grained radio control decisions through xApps interacting with O-DU and O-RU nodes via the E2 interface \cite{O-RanE2Interface}. This hierarchical control paradigm naturally supports the integration of sustainability objectives into operational decision-making: long-term carbon-reduction policies can be translated into near-real-time resource management actions. As a result, O-RAN evolves from a flexible architectural framework into a programmable platform capable of supporting intelligent, carbon-aware radio access network operation.

\subsection{Related Work}
\ac{BS} sleeping\cite{liang2026scalable} has been extensively studied as a means to reduce energy consumption in cellular networks, especially during off-peak hours. Wu \textit{et al.}~\cite{wu2013traffic} proposed two traffic-aware sleep strategies, threshold-based and vacation-based, that jointly optimize BS activation and transmit power, revealing a fundamental tradeoff between energy savings and user delay. 
In heterogeneous networks, energy savings can be achieved by selectively deactivating underutilized \acp{BS}. Zhou \textit{et al.}~\cite{zhou2023joint} formulated a joint user association and \ac{BS} operation framework to optimize long-term network utility under energy constraints. Their results show that activating only a subset of \acp{SBS} based on traffic demand can significantly reduce energy use without degrading service quality.
To incorporate renewable energy sources, Gong \textit{et al.}~\cite{gong2014base} considered a two-stage optimization scheme for \ac{BS} sleeping and subcarrier allocation, aiming to minimize non-renewable power consumption. Their approach dynamically adjusts \ac{BS} states based on both harvested energy availability and user traffic, striking a balance between energy cost and blocking probability.
Recent architectural advances enable finer-grained sleep control. Sun \textit{et al.}~\cite{sun2025flexible} investigated \ac{BS} sleeping in uplink--downlink fully decoupled RANs, jointly optimizing user association, power allocation, and BS activity. Similarly, Wu \textit{et al.}~\cite{wu2021smdp} modeled \ac{BS} sleep behavior using a \ac{SMDP}, enabling dynamic sleep scheduling in macro-femto networks based on user arrival patterns and energy efficiency.
In addition to optimization-based methods, traffic prediction has been used to guide sleep decisions. Lin \textit{et al.}~\cite{lin2021data} proposed a data-driven \ac{BS} sleeping strategy based on spatio-temporal traffic forecasting using \ac{GNN} and LSTMs. Spatio--temporal correlations are further leveraged in~\cite{wu2021deep}, where convolutional-\ac{LSTM} forecasting is coupled with an actor--critic controller to improve robustness under fluctuating traffic.

Recent studies have increasingly adopted \ac{DRL}-based paradigms to tackle the \ac{BS} sleep/active control problem, motivated by their ability to handle high-dimensional state spaces, long-term dependencies, and non-linear decision boundaries. 
Several representative learning-based solutions have been proposed in this context. For instance, \cite{ju2022energy}, the authors propose a \ac{DQN}-based framework augmented with an auxiliary action-selection network that filters infeasible actions, thereby improving exploration efficiency and stabilizing training.
To explicitly account for switching overheads and service degradation, \cite{ye2019drag} introduces a modular \ac{DRL} architecture that combines a \ac{DDPG} policy with a supervised cost estimator and a traffic prediction module. This hybrid design allows the controller to internalize the long-term impact of sleep transitions. In \cite{sun2024deep}, a \ac{PPO}-based solution is developed to simultaneously coordinate sleep scheduling, cell zooming, user association, and \ac{RIS} configuration, addressing the coupling between discrete and continuous control variables in dense multi-cell environments.
Risk-awareness and robustness have also received growing attention. In \cite{masoudi2022digital}, a digital twin is integrated with a \ac{DQN} controller to evaluate delay risk prior to action execution, enabling online retraining or feature suppression under abnormal traffic conditions. From a distributed control perspective, \cite{zhen2024energy} models each \ac{BS} as an autonomous agent and applies a multi-agent \ac{PPO} with a state-similarity mechanism, jointly optimizing BS sleeping and MIMO antenna activation while mitigating oscillatory behavior.

Beyond pure energy minimization, recent green RAN research increasingly emphasizes where the energy is drawn from and how operational decisions translate into carbon footprint. In renewable energy powered cellular networks, the key objective is often reframed from minimizing total energy to minimizing on-grid energy, since renewable consumption is typically treated as carbon-free and its availability is both stochastic and spatially imbalanced across sites. For example, the paper \cite{lee2019adaptive} studies \acp{BS} powered by hybrid supplies and jointly optimize traffic management and energy cooperation to reduce grid energy usage while meeting throughput constraints, highlighting that renewable-aware topology/traffic control and energy exchange between sites are complementary levers rather than substitutes.
Building on this viewpoint, a line of work investigates joint sleep control and renewable energy sharing in dense deployments. \cite{gan2025joint} formulate \ac{SBS} sleep decisions together with renewable energy sharing as a unified \ac{MDP}, aiming to reduce grid energy consumption and improve renewable utilization under time-varying traffic and harvested energy. To address the curse of dimensionality in \ac{UDN}-scale deployments, they adopt a multi-discrete PPO design that factorizes the combinatorial control space into tractable sub-actions, demonstrating notable gains over schemes that treat sleeping and energy sharing separately. Complementary to terrestrial-only designs, vertical-assisted architectures have also been explored as a sustainability enabler. Paper~\cite{salamatmoghadasi2023energy} leverage a \ac{HAPS} layer to offload traffic and enable more aggressive \ac{SBS} sleeping, reporting sizable reductions in grid power consumption while maintaining \ac{QoS}.
Another strand explicitly models carbon footprint rather than energy alone and shows that operational policies interact with both operational and embodied emissions. \cite{renga2025network} analyze network sharing as a mechanism to enable recurrent BS deactivation during low-load periods and quantify carbon-footprint mitigation beyond energy savings, including the effect of switching frequency on equipment degradation and embodied carbon through lifetime extension. From an architectural perspective, O-RAN sustainability studies provide foundational power-consumption models and highlight that open and cloud-native O-RAN implementations may shift energy burdens across \ac{RU}, \ac{DU} and \ac{CU} and transport components, motivating careful design of energy-aware control loops. For example, Baldini \emph{et al.}~\cite{baldini2024toward} develop a parametric power model for O-RAN and use real RU deployment datasets to expose how placement and processing/transport assumptions shape end-to-end power trends, thus informing where carbon-aware operation is most impactful.

\subsection{Contribution}
Although the above studies are closely related to sustainable wireless network operation, they differ fundamentally from the focus of this paper. In this work, we do not consider energy sharing, battery dynamics, or joint traffic management. Instead,
In this work, we consider an \ac{O-RAN} system where \acp{RU} are powered by heterogeneous energy sources, including renewable energy and the conventional power grid. By explicitly distinguishing renewable-powered \acp{RU}, which incur zero operational carbon emissions, from grid-powered \acp{RU}, we reformulate the \ac{RU} control problem as a carbon-aware sequential decision making task. This formulation fundamentally alters the structure of the control problem: RU activation decisions are no longer symmetric, and the same action may lead to different carbon outcomes depending on the underlying energy source. As a result, conventional energy-centric control policies may become suboptimal when carbon reduction is a primary objective.
Moreover, unlike \cite{gan2025joint}, which develops a customized multi-discrete PPO framework for large-scale joint control, we conduct a systematic comparison between deterministic off-policy learning and stochastic on-policy learning under the same carbon-aware MDP formulation. This enables us to reveal how different DRL paradigms respond to reward shaping and heterogeneous energy supply in sustainable O-RAN operation. Through extensive simulations, we compare representative policy-gradient and actor--critic methods and analyze their learning behaviors under the proposed sustainability-aware formulation.

The main contributions of this paper are summarized as follows:

\begin{itemize}
    \item We formulate a carbon-aware \ac{RU} sleep mode control problem by explicitly modeling heterogeneous energy sources, where renewable-powered and grid-powered \acp{RU} lead to different carbon implications under identical activation decisions.

    \item We design a sustainability-oriented reward function within the \ac{MDP} framework that jointly captures network energy consumption, operational carbon emissions, and QoS satisfaction, enabling the learning agent to balance efficiency and environmental impact.

    \item We implement and evaluate representative \ac{DRL} approaches, including deterministic off-policy (TD3) and stochastic on-policy (PPO) methods, under a unified simulation framework, ensuring fair comparison across learning paradigms.

    \item We provide a comprehensive performance analysis across multiple deployment scenarios, demonstrating that carbon-aware optimization introduces a trade-off between energy efficiency and emissions, and showing that TD3 achieves more stable convergence and consistently better energy--carbon performance compared to PPO and heuristic model.
\end{itemize}

\section{System Model}

\begin{figure}[!t]
	\centering
	\includegraphics[clip, trim=0.0cm 0.cm 0.0cm 0cm, width=1.0\columnwidth]{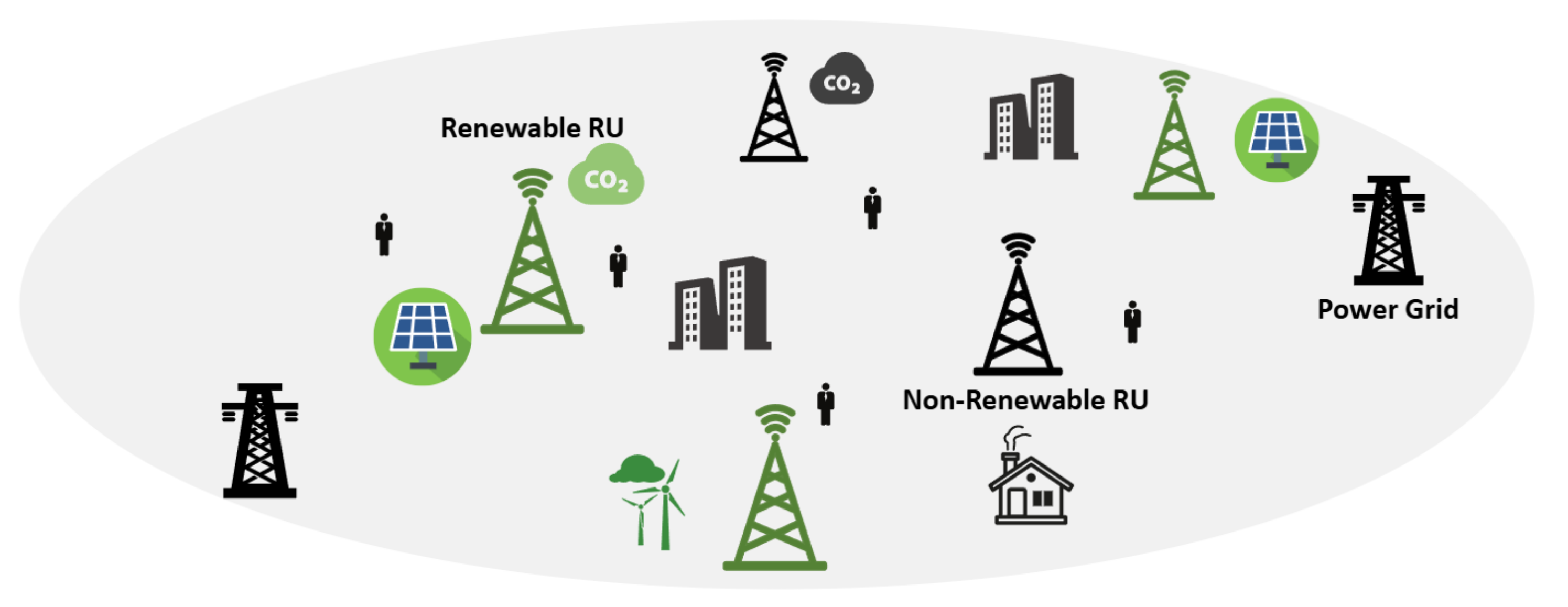}
	\caption{Proposed heterogeneous energy network }
	\label{fig:systemmodel1}
\end{figure}

We consider an \ac{O-RAN}-based cellular network comprising a fixed set of \acp{RU}, denoted by $\mathcal{M}=\{1,\ldots,M\}$, and a set of \acp{UE}, denoted by $\mathcal{K}=\{1,\ldots,K\}$. The RU positions are predetermined and remain unchanged, as shown in Fig.~\ref{fig:systemmodel1}. Time is divided into $T$ slots indexed by $t\in\{1,\ldots,T\}$.

To model heterogeneous electricity-procurement arrangements, the set of RUs is partitioned into two disjoint subsets: (i) renewable-matched RUs, denoted by $\mathcal{M}_{\mathrm{R}}$, whose electricity demand is contractually matched with renewable generation, and (ii) non-renewable-supplied RUs, denoted by $\mathcal{M}_{\mathrm{G}}$, whose electricity demand is assigned the emission factor of the residual non-renewable grid supply. These subsets satisfy
\begin{equation}
\mathcal{M}_{\mathrm{R}}\cup\mathcal{M}_{\mathrm{G}}
=\mathcal{M},
\qquad
\mathcal{M}_{\mathrm{R}}\cap\mathcal{M}_{\mathrm{G}}
=\varnothing.
\end{equation}

Unlike deployments with on-site renewable generation, we adopt a source-matched, market-based operational accounting model. Electricity consumed by both RU categories may be physically delivered through the same power grid; therefore, the partition does not represent physically separated electricity flows. Instead, it represents different contractual generation attributes assigned to the electricity consumed by the RUs.

The renewable-energy contracts are assumed to provide sufficient matching for the electricity demand of RUs in $\mathcal{M}_{\mathrm{R}}$ over the considered control horizon. This is an exogenous scenario assumption rather than a decision made by the DRL agent. Renewable-generation forecasting, electricity procurement, storage operation, curtailment, and energy sharing are outside the scope of this work.

Under the adopted operational accounting boundary, RUs in $\mathcal{M}_{\mathrm{R}}$ are assigned zero direct operational emissions at the point of renewable-electricity generation, whereas RUs in $\mathcal{M}_{\mathrm{G}}$ are assigned a positive source-specific operational emission factor. This boundary does not include embodied or life-cycle emissions associated with energy-generation infrastructure. The resulting abstraction enables us to isolate the effect of heterogeneous electricity-procurement arrangements on carbon-aware RU activation control.

Each RU $m \in \mathcal{M}$ is equipped with $Q_m$ \acp{PRB}. Denote by $n_{m,k}^t$ the number of \acp{PRB} allocated from RU $m$ to UE $k$ at time slot $t$. The total \ac{PRB} usage of RU $m$ is
\begin{equation}\label{eq: ru_load}
N_m^t = \sum_{k \in \mathcal{K}} n_{m,k}^t, 
\qquad l_m^t = \frac{N_m^t}{Q_m},
\end{equation}
where $l_m^t$ is the normalized \ac{PRB} load. The \ac{UE} association and \ac{PRB} allocation are handled by the simulation environment. Initially, each UE is associated with its nearest RU. When the serving RU is switched off, the UE is reassigned to the nearest active RU. During UE mobility, a handover to another active candidate RU is performed when the candidate requires fewer PRBs to meet the target rate and has sufficient residual PRB capacity.
For each UE, the requested number of \acp{PRB} is calculated as the minimum number required to achieve $R_k^{\min}$ under the current channel condition. If the total number of requested PRBs does not exceed the capacity $Q_m$ of RU $m$, each UE receives its requested PRBs. Otherwise, the available PRBs are allocated proportionally among the UEs associated with that RU while satisfying the RU-level PRB capacity. The achieved UE data rates are subsequently recalculated based on the allocated PRBs.

The \ac{SNR} of \ac{UE} $k$ served by RU $m$ at slot $t$ is modeled as
\begin{equation}
\gamma_{m,k}^t = \frac{P_{\mathrm{TX}} g_{m,k}^t}{Q_mN_0 W_{\mathrm{PRB}}},
\end{equation}
where $P_{\mathrm{TX}}$ is the transmission power, $g_{m,k}^t$ is the gain of the channel including path loss and fading, $N_0$ is the spectral density of the noise power and $W_{\mathrm{PRB}}$ is the bandwidth of a \ac{PRB}. The achievable data rate of \ac{UE} $k$ is then
\begin{equation}
R_{k}^t = \sum_{m \in \mathcal{M}} n_{m,k}^t \, W_{\mathrm{PRB}} 
\log_2\!\left(1+\gamma_{m,k}^t\right).
\end{equation}
To guarantee service quality, each \ac{UE} must satisfy a minimum data rate requirement.

\subsection{Power Consumption Model}

The activation status of \ac{RU} $m$ is described by a binary variable $\alpha_m^t \in \{0,1\}$, where $\alpha_m^t=1$ if RU $m$ is active and $\alpha_m^t=0$ if \ac{RU} $m$ is in sleep mode. The power consumption of each \ac{RU} is composed of three components~\cite{ju2022energy}. The first is fixed power consumption, denoted by $P_{m}^{\mathrm{Fix},t}$, which represents the energy used by signal processing, cooling systems, and power supply units. This component depends solely on the operational mode of the \ac{RU}. 
The second component is the load-dependent power consumption, denoted by $P_{m}^{\mathrm{data},t}$, primarily attributed to the \ac{PA}. In this work, the load is quantified based on the \ac{PRB} utilization, as defined in~\eqref{eq: ru_load}. The third component is the transition power $P_{m}^{\mathrm{trans},t}$, which is incurred only when the \ac{RU} switches from sleep mode to active mode. By summing these three components, the total power consumption of \ac{RU} $m$ at time slot $t$ is given by:
\begin{equation}
P_m^t = P_{\mathrm{fix},m}^t + P_{\mathrm{data},m}^t + P_{\mathrm{trans},m}^t,
\end{equation}
with
\begin{align}
P_{\mathrm{fix},m}^t &= \alpha_m^t P_m^{\mathrm{active}} 
+ (1-\alpha_m^t) P_m^{\mathrm{sleep}}, \\
P_{\mathrm{data},m}^t &= \alpha_m^t \frac{P_{\mathrm{TX}}}{\eta}\, l_m^t, \\
P_{\mathrm{trans},m}^t &= \max\left( \alpha_m^t - \alpha_m^{t-1}, 0 \right) \cdot V_m^{trans}.
\end{align}
Here $P_m^{\mathrm{active}}$ and $P_m^{\mathrm{sleep}}$ are the baseline power levels in active and sleep states, respectively; $\eta$ is the power amplifier efficiency; and $V_m^{\mathrm{trans}}$ captures the overhead due to switching from sleep to active states.

\subsection{Carbon Emission Model}

We consider direct operational emissions associated with the electricity consumed by the RUs. Let $c_m^t$ [kgCO$_2$/kWh] denote the source-specific operational carbon intensity assigned to RU $m$ at time slot $t$ \cite{DING2022106339}. Under the adopted market-based accounting boundary, the carbon intensity used in the simulations is
\begin{equation}
\label{eq:carbon-intensity-class}
c_m^t =
\begin{cases}
0, & m\in\mathcal{M}_{\mathrm{R}},\\
c_{\mathrm{G}}, & m\in\mathcal{M}_{\mathrm{G}},
\end{cases}
\end{equation}
where $c_{\mathrm{G}}>0$ is the operational carbon intensity assigned to the non-renewable electricity supply. The zero value assigned to renewable-matched RUs represents zero direct emissions at the point of renewable-electricity generation. It does not include embodied or life-cycle emissions associated with energy-generation infrastructure.

Since $P_m^t$ is measured in watts and the slot duration $\Delta t$ is measured in seconds, the electricity consumed by RU $m$ during time slot $t$ is $E_m^t = \frac{\Delta t}{3.6\times10^{6}}P_m^t
\quad [\mathrm{kWh}].$
Accordingly, the operational carbon emissions of RU $m$ are
\begin{equation}
\label{eq:co2-per-ru}
\mathrm{CO}_{2,m}^t
=
E_m^t c_m^t
=
\frac{\Delta t}{3.6\times10^{6}}c_m^t P_m^t
\quad [\mathrm{kgCO}_2].
\end{equation}

The network-wide operational carbon emissions during time slot $t$ are
\begin{equation}
\label{eq:co2-network}
\mathrm{CO}_{2}^t
=
\sum_{m\in\mathcal{M}}\mathrm{CO}_{2,m}^t
=
\frac{\Delta t}{3.6\times10^{6}}
\sum_{m\in\mathcal{M}}c_m^tP_m^t.
\end{equation}
The cumulative operational emissions over the horizon $\{1,\ldots,T\}$ are
\begin{equation}
\label{eq:co2-cumulative}
\mathrm{CO}_{2,\mathrm{tot}}
=
\sum_{t=1}^{T}\mathrm{CO}_{2}^t
=
\frac{\Delta t}{3.6\times10^{6}}
\sum_{t=1}^{T}\sum_{m\in\mathcal{M}}c_m^tP_m^t.
\end{equation}

The notation $c_m^t$ retains both RU and time indices so that the general formulation can accommodate RU-specific or time-varying carbon-intensity inputs. In the present simulations, however, we use the two source-specific values in~\eqref{eq:carbon-intensity-class} to isolate the effect of heterogeneous electricity procurement on RU activation control.



\subsection{Problem Formulation}

Given the system model, the network operation optimizes RU activity $\{\alpha_m^t\}$ over slots $t=1,\dots,T$. The objective is to minimize a weighted sum of renewable and non-renewable energy consumption:
\begin{align}
&\min_{\{\alpha_m^t\}}\ \sum_{t=1}^{T}
\begin{aligned}[t]
\Bigl(&\lambda \sum_{m\in\mathcal{M}_{\mathrm{R}}} P_m^t\\
&{}+(1-\lambda)\sum_{m\in\mathcal{M}_{\mathrm{G}}}P_m^t\Bigr),
\end{aligned}\label{eq:pf_obj}\\
\text{s.t.}\quad
& R_k^t \;\ge\; R_k^{\min}, \quad \forall k,t, \label{c:qos}\\
& \sum_{k\in\mathcal{K}} n_{m,k}^t \;\le\; Q_m, \quad \forall m,t, \label{c:prb}\\
& \sum_{m\in\mathcal{M}} \delta_{m,k}^t = 1, \quad \forall k,t, \label{c:assoc}\\
&\begin{aligned}[t]
n_{m,k}^t&\in\mathbb{Z}_{0},\quad
\alpha_m^t,\delta_{m,k}^t\in\{0,1\},\\
&\hspace{7em}\forall m,k,t.
\end{aligned}\label{c:int}
\end{align}

\noindent
Here $\mathcal{M}_{\mathrm{R}}$ and $\mathcal{M}_{\mathrm{G}}$ denote the sets of renewable-energy and non-renewable-energy RUs, respectively. $P_m^t$ is the instantaneous power consumption of RU $m$ given its operational state and traffic load. The scalar $\lambda \in [0,1]$ controls the relative emphasis between renewable and non-renewable energy consumption. $R_k^t$ is the downlink rate of UE $k$ at slot $t$ and must satisfy the QoS requirement $R_k^{\min}$. Constraint \eqref{c:prb} limits the PRB allocation per RU, while \eqref{c:assoc} enforces unique RU association for each UE. 

\smallskip
\noindent
Note that explicit carbon-emission equations are omitted from the formulation. Since only non-renewable consumption contributes to carbon emissions, minimizing the weighted energy consumption in \eqref{eq:pf_obj} naturally encourages deactivation of non-renewable RUs, thereby reducing the carbon footprint.

\section{Markov Decision Process Formulation}
In this section, we formulate the \ac{RU} sleep/active control problem as a \ac{MDP}, which enables sequential decision-making under dynamic traffic, user mobility, and heterogeneous energy sources. At each time slot, the agent observes the network state, selects a joint \ac{RU} activation action, and receives a reward that reflects the sustainability and service performance of the network. The environment then evolves according to user mobility, traffic demand variations, and wireless channel dynamics. This formulation allows the agent to learn long-term policies that balance energy efficiency, carbon emission reduction, and \ac{QoS} satisfaction over time.

Formally, the MDP is defined by the tuple $(\mathcal{S}, \mathcal{A}, \mathcal{P}, r)$, where $\mathcal{S}$ denotes the state space, $\mathcal{A}$ the action space, $\mathcal{P}$ the state transition probability, and $r$ the reward function. The detailed definitions of the state space, action space, and reward function are given as follows.

\subsubsection{State Space}
The state comprises essential information required for policy learning. At each time step $t$, the state includes the real data rates of all \acp{UE}, expressed as
\begin{equation}
    \mathbf{R}_{d}^{t} = \begin{bmatrix} R_{d,1}^{t}, R_{d,2}^{t}, \dots, R_{d,K}^{t} \end{bmatrix}^{T},
\end{equation}
which reflects the system's ability to meet user data rate requirements. The activation states of the \acp{RU} from the previous time step are given by
\begin{equation}
    \boldsymbol{\alpha}^{t-1} = \begin{bmatrix} \alpha_{1}^{t-1}, \alpha_{2}^{t-1}, \dots, \alpha_{M}^{t-1} \end{bmatrix}^{T}.
\end{equation}
This information captures the temporal coupling between successive RU activation decisions and allows the agent to account for switching behavior.

The \ac{PRB} utilization levels of the RUs are represented as
\begin{equation}
    \mathbf{L}^{t} = \begin{bmatrix} l_{1}^{t}, l_{2}^{t}, \dots, l_{M}^{t} \end{bmatrix}^{T},
\end{equation}
which quantify the instantaneous traffic load of each RU. In addition, the spatial positions of the \acp{UE} are included as
\begin{equation}
\mathbf{U}^{t}=\bigl[(x_{1}^{t},y_{1}^{t}),(x_{2}^{t},y_{2}^{t}),\ldots, (x_{K}^{t},y_{K}^{t})\bigr]^{T},
\end{equation}
where $(x_k^{t}, y_k^{t})$ denote the coordinates of UE $k$ at time $t$. The UE positions directly affect path loss, user association, and achievable data rates, and thus influence RU load distributions. Including spatial information in the state enables the agent to anticipate future handovers, traffic shifts, and coverage-critical regions, rather than reacting solely to instantaneous load conditions.

Overall, the state at time $t$ is defined as
\begin{equation}
    s_t = \begin{bmatrix} \mathbf{R}_{d}^{t}, \boldsymbol{\alpha}^{t-1}, \mathbf{L}^{t}, \mathbf{U}^{t} \end{bmatrix}^{T}.
\end{equation}
All state variables are normalized to comparable ranges to facilitate stable and efficient learning.

\subsubsection{Action Space}
The action $a_t \in \mathcal{A}$ represents the operational states of all \acp{RU} and is defined as
\begin{equation}
    \mathcal{A} = \left[ \alpha_1, \alpha_2, \dots, \alpha_M \right],
\end{equation}
where each element $\alpha_m \in \{0,1\}$ denotes the operating mode of RU $m$. Specifically, $\alpha_m = 1$ indicates that RU $m$ is active, while $\alpha_m = 0$ corresponds to sleep mode. The action is generated by the \ac{DRL} policy and determines the network-wide RU activation pattern at each time slot.

\subsubsection{Reward Function}
The reward function is designed to guide the agent toward sustainability-aware RU control decisions by balancing energy efficiency, carbon emission reduction, and \ac{QoS} satisfaction. To evaluate different sustainability objectives, we consider two carbon-aware reward formulations.

a) Energy--Carbon--QoS reward:
\begin{equation}
r_t^{(1)}=-w_1\cdot\text{norm}\!\left(P_{\text{tot}}^{t}\right)
-w_2\cdot\text{norm}\!\left(C_{\text{tot}}^{t}\right)
-w_3\cdot\text{norm}\!\left(K_{\text{unsat}}^{t}\right),
\end{equation}
where $P_{\text{tot}}^{t}$ denotes the total network energy consumption at time $t$, $C_{\text{tot}}^{t}$ represents the corresponding carbon emissions, and $K_{\text{unsat}}^{t}$ is the number of UEs whose data rates fall below the minimum QoS requirement. Each term is normalized to ensure comparable scales and to prevent dominance by any single component. This reward formulation jointly optimizes energy efficiency, carbon reduction, and service reliability.

b) Carbon--QoS-oriented reward:
\begin{equation}
r_t^{(2)}=-w_4\cdot\text{norm}\!\left(C_{\text{tot}}^{t}\right)
-w_5\cdot\text{norm}\!\left(K_{\text{unsat}}^{t}\right),
\end{equation}
which focuses explicitly on minimizing operational carbon emissions while maintaining QoS constraints. By removing the explicit energy term, this reward highlights scenarios where carbon reduction, rather than energy minimization, is the dominant objective, particularly in networks with heterogeneous renewable-powered and grid-powered RUs. The parameter \(\lambda\) in Eq.\eqref{eq:pf_obj} provides a high-level representation of the relative importance of renewable and non-renewable energy consumption. This trade-off is implemented in the MDP through the reward weights. AG1 penalizes both total energy consumption and carbon emissions, AG2 penalizes only carbon emissions and therefore places the strongest preference on renewable-matched operation. Thus, the two reward formulations correspond to different implementations of the source-weighting principle represented by \(\lambda\). And in both reward formulations, QoS constraints $R_{d,k}^{t} \ge R_{\text{min},d,k}$ are enforced through the penalty term associated with $K_{\text{unsat}}^{t}$. The temporal accumulation of this penalty discourages persistent QoS violations and promotes stable network operation. 

\begin{figure}[!t]
	\centering
	\includegraphics[clip, trim=0.0cm 0.cm 0.0cm 0cm, width=1.0\columnwidth]{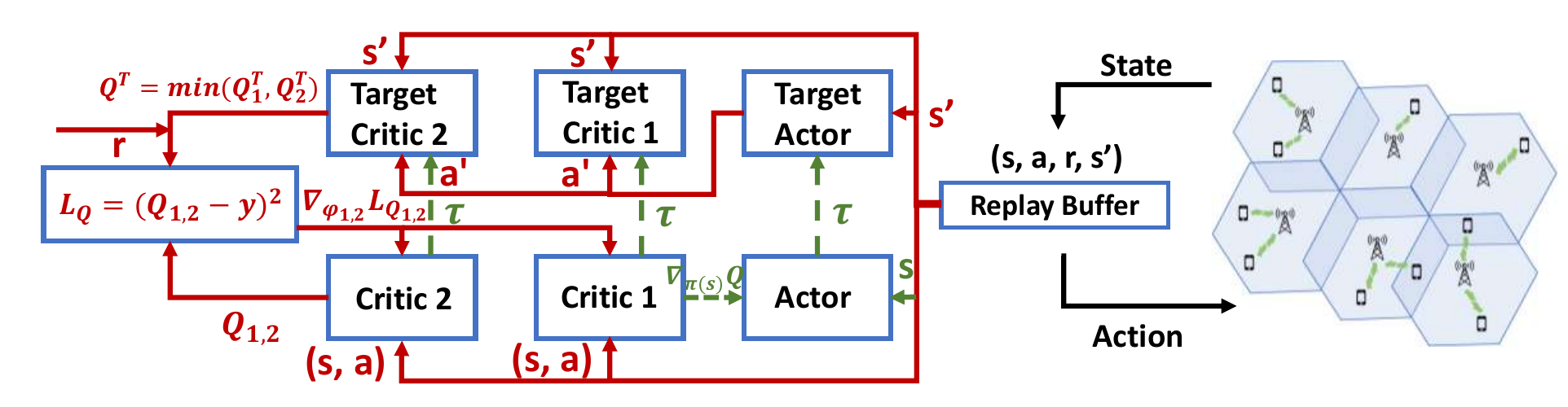}
	\caption{Illustrates the workflow of TD3 training process. Red solid lines indicate critic training based on temporal-difference loss; green dashed lines represent actor updates guided by critic gradients and soft update target network. }
	\label{fig:systemmodel}
\end{figure}

\section{Sustainable DRL-based RU sleep control}

To solve the formulated MDP, we adopt \ac{DRL} methods to learn optimal RU activation policies under dynamic traffic variations and heterogeneous energy conditions. In particular, we consider two representative DRL paradigms: deterministic off-policy learning and stochastic on-policy learning. 

Off-policy methods, such as TD3, learn the optimal policy using transitions sampled from a replay buffer, which may contain experiences generated by previous policies. This allows for improved sample efficiency and stable learning, as the agent can reuse past experiences and decorrelate data samples. Such properties are particularly advantageous in our setting, where the RU activation problem exhibits a high-dimensional combinatorial action space and strong coupling between decisions across RUs. The ability to leverage historical interactions enables more effective exploration of the action space and refinement of activation patterns over time.

In contrast, on-policy methods, such as PPO, update the policy using only data collected from the current policy, ensuring that the learning process closely follows the latest policy behavior. This results in more stable and theoretically grounded updates, as the policy gradient is computed with respect to the current data distribution. However, this also leads to lower sample efficiency and may limit exploration in complex environments. In the context of RU sleep control, where the reward landscape can be highly non-convex due to QoS constraints and heterogeneous energy sources, on-policy methods may converge faster in early stages but risk settling into suboptimal solutions.

Overall, the comparison between off-policy and on-policy learning provides valuable insights into the trade-offs between exploration, stability, and convergence performance in sustainability-aware RU control. This design enables a systematic evaluation of how different learning paradigms respond to the structural complexity of the problem.

\subsection{TD3-Based Training for RU Sleep Control}

The joint sleep/active control of multiple \acp{RU} inherently leads to a high-dimensional and strongly coupled action space. When the number of controllable RUs increases, discrete value-based approaches face an exponential growth of possible action combinations, which severely degrades exploration efficiency and learning stability. To overcome this limitation, we employ a continuous-control formulation based on the \ac{TD3} algorithm, which enables learning by relaxing the discrete RU activation decisions into a continuous action space during training.

\subsubsection{Continuous Relaxation of RU Activation}
In the proposed framework, the actor network parameterized by $\theta$ outputs a continuous action vector
\begin{equation}
\boldsymbol{\mu}_{\theta}(s)=\bigl[\mu_{\theta,1}(s),\mu_{\theta,2}(s),\ldots,\\
\mu_{\theta,M}(s)\bigr]\in[0,1]^M,
\end{equation}
where each element represents the activation tendency of an RU. This continuous representation allows the policy to express fine-grained preferences over RU activation levels, thereby avoiding direct enumeration of all binary activation patterns. Note that, the continuous actor output is interpreted as a proto-action rather than the physical operating state of an RU. Binary actions are executed in the environment and used to train the critics. During the actor update, the critic is evaluated at the continuous proto-action and provides a surrogate gradient over the relaxed action space. This approximation enables deterministic policy-gradient training without explicitly enumerating all joint binary actions. However, the resulting gradient is not the exact gradient of the thresholded binary policy, and approximation errors may arise between the interior of the relaxed action space and its binary vertices.

During interaction with the environment, the continuous actions are converted into executable binary decisions through a deterministic discretization mechanism combined with exploration noise. Specifically, Gaussian noise is added to each action dimension to encourage exploration,
\begin{equation}
    \tilde{a}_i = \mu_{\theta,i}(s) + \eta_i, \quad \eta_i \sim \mathcal{N}(0, \sigma^2),
\end{equation}
followed by threshold-based discretization;
\begin{equation}
    a_i =
    \begin{cases}
        1, & \text{if } \tilde{a}_i > 0.5, \\
        0, & \text{otherwise},
    \end{cases}
\end{equation}
where $a_i$ denotes the operational state of RU $i$. This hybrid continuous--discrete execution strategy enables efficient exploration in a smooth action space while ensuring compatibility with binary RU sleep/active control.

\subsubsection{Critic Learning and Target Estimation}
The learning process relies on an experience replay buffer that stores transitions $(s_t, a_t, r_t, s_{t+1})$, where the reward reflects energy consumption, carbon emissions, and QoS satisfaction according to the adopted sustainability objective.

To mitigate overestimation bias in value approximation, TD3 employs two independent critic networks, $Q_{\phi_1}$ and $Q_{\phi_2}$, both of which estimate the state--action value function.

For each sampled transition, the target Q-value is computed using the minimum of the two target critics,
\begin{equation}
y_t=r_t+\gamma\min_{i\in\{1,2\}} Q_{\phi_i'}\!\Bigl(s_{t+1}, f_d\!\left(\boldsymbol{\mu}_{\theta'}(s_{t+1})+\epsilon\right)\Bigr),
\end{equation}
where $\theta'$ and $\phi_i'$ denote the parameters of the target actor and target critics, respectively. The noise term $\epsilon \sim \text{clip}(\mathcal{N}(0, \sigma_{\text{target}}^2))$ is used for target policy smoothing, which reduces sensitivity to small perturbations in the action space and improves learning robustness.

Each critic network is updated by minimizing the mean squared temporal-difference loss
\begin{equation}
    \mathcal{L}(\phi_i) = \mathbb{E}_{(s,a,r,s') \sim \mathcal{D}}
    \left[ \left(Q_{\phi_i}(s,a) - y_t \right)^2 \right].
\end{equation}

\subsubsection{Actor Update and Delayed Policy Optimization}
The actor network is updated using the deterministic policy gradient, which adjusts the policy parameters to maximize the expected Q-value estimated by the critics,
\begin{equation}
\nabla_{\theta}J(\theta)=\mathbb{E}_{s\sim\mathcal{D}}\Bigl[
\nabla_a Q_{\phi_1}(s,a)\big|_{a=\boldsymbol{\mu}_{\theta}(s)}\cdot\nabla_{\theta}\boldsymbol{\mu}_{\theta}(s)\Bigr].
\end{equation}
To enhance stability, actor updates are performed less frequently than critic updates, following the delayed policy update strategy of TD3. This separation allows the critics to provide more accurate value estimates before the policy is adjusted.

\subsubsection{Target Network Updates}
Both the actor and critic target networks are updated using soft updates,
\begin{equation}
\theta'\leftarrow\tau\theta+(1-\tau)\theta',\quad \phi_i'\leftarrow\tau\phi_i+(1-\tau)\phi_i',\quad i\in\{1,2\},
\end{equation}
where $\tau \ll 1$ controls the update rate. This mechanism smooths parameter evolution and further stabilizes training.

\begin{algorithm}[t]
\small
\caption{TD3 Training for Sustainability-Aware RU Sleep Control}
\label{alg:td3}
\KwIn{Initial state $s_0$; actor $\mu_{\theta}$; critics $Q_{\phi_1}, Q_{\phi_2}$; replay buffer $\mathcal{D}$; discount $\gamma$; noise scales $\sigma,\sigma_{\text{tgt}}$; update delay $d$; soft update rate $\tau$}
\KwOut{Trained actor policy $\mu_{\theta}$ for joint RU activation control}

\For{episode $=1,2,\dots$}{
    Reset environment and observe $s \leftarrow s_0$\;
    \For{$t=1,2,\dots,T$}{
        \tcp{(1) Continuous activation scores for all RUs}
        Draw exploration noise $\boldsymbol{\eta}\sim\mathcal{N}(\mathbf{0},\sigma^2\mathbf{I})$\;
        Compute activation scores $\mathbf{z} \leftarrow \mu_{\theta}(s) + \boldsymbol{\eta}$, where $\mathbf{z}\in\mathbb{R}^M$\;
        Clip $\mathbf{z}$ to $[0,1]^M$\;

        \tcp{(2) Discretize to executable RU on/off decisions}
        Construct RU action vector $\mathbf{a}\in\{0,1\}^M$ by
        $a_m \leftarrow \mathbb{I}\!\left[z_m>\delta\right],~\forall m\in\{1,\dots,M\}$\;
        \tcp{Optionally: $\delta=0.5$; interpret $\{m:a_m=1\}$ as the selected active RU set}

        \tcp{(3) Environment interaction}
        Execute $\mathbf{a}$, obtain reward $r$ and next state $s'$ (reward uses the chosen carbon-aware objective)\;
        Store transition $(s,\mathbf{a},r,s')$ into $\mathcal{D}$\;
        Set $s \leftarrow s'$\;

        \tcp{(4) Critic update}
        Sample a mini-batch $\{(s_j,\mathbf{a}_j,r_j,s'_j)\}_{j=1}^{N}$ from $\mathcal{D}$\;
        Draw target smoothing noise $\boldsymbol{\epsilon}\sim\text{clip}(\mathcal{N}(\mathbf{0},\sigma_{\text{tgt}}^2\mathbf{I}))$\;
        Compute target scores $\mathbf{z}'_j \leftarrow \mu_{\theta'}(s'_j)+\boldsymbol{\epsilon}$, clip to $[0,1]^M$\;
        Discretize target action $\tilde{\mathbf{a}}'_j$ by $\tilde{a}'_{j,m}\leftarrow \mathbb{I}[z'_{j,m}>\delta]$\;
        Compute TD target:
        $y_j \leftarrow r_j + \gamma \min_{i\in\{1,2\}} Q_{\phi_i'}(s'_j,\tilde{\mathbf{a}}'_j)$\;
        Update critics by minimizing, for $i\in\{1,2\}$:
        $\mathcal{L}(\phi_i)=\frac{1}{N}\sum_{j=1}^{N}\left(Q_{\phi_i}(s_j,\mathbf{a}_j)-y_j\right)^2$\;

        \tcp{(5) Delayed actor and target updates}
        \If(\tcp*[f]{update every $d$ steps}){$t \bmod d = 0$}{
            Update actor using deterministic policy gradient:\;
            \hspace{0.6em}$\nabla_{\theta}J \approx \frac{1}{N}\sum_{j=1}^{N} 
            \nabla_{\mathbf{a}} Q_{\phi_1}(s_j,\mathbf{a})\big|_{\mathbf{a}=\mu_{\theta}(s_j)}
            \nabla_{\theta}\mu_{\theta}(s_j)$\;

            Soft update target networks:\;
            \hspace{0.6em}$\theta' \leftarrow \tau \theta + (1-\tau)\theta'$\;
            \hspace{0.6em}$\phi_i' \leftarrow \tau \phi_i + (1-\tau)\phi_i',~~i\in\{1,2\}$\;
        }
    }
}
\end{algorithm}

\subsection{PPO-Based Training for RU Sleep Control}

To provide a representative stochastic policy-gradient baseline, we implement \ac{PPO} for the same RU sleep/active control task and environment as the TD3 agent. In our setting, the controller must decide, at every time slot, whether each RU should be active or sleeping, resulting in a multi-dimensional binary action vector $\mathbf{a}_t \in \{0,1\}^{M}$. Unlike TD3, which learns a deterministic continuous policy and applies thresholding for execution, PPO directly models the joint RU activation decision as a stochastic policy over binary actions. This design enables on-policy learning with clipped policy updates while preserving the discrete nature of RU switching.

\subsubsection{Policy Parameterization for Multi-RU Binary Actions}
We adopt an actor-critic architecture with a shared feature extractor and two heads: an actor head producing RU-wise logits and a critic head estimating the state value. Given the observed state $s_t$, the network outputs
\begin{equation}
    \boldsymbol{\ell}_t,\, V_{\psi}(s_t) = f_{\psi}(s_t),
\end{equation}
where $\boldsymbol{\ell}_t \in \mathbb{R}^{M}$ denotes the logits for all RUs, and $V_{\psi}(s_t)$ is the state-value estimate. The RU activation policy is modeled as a factorized Bernoulli distribution
\begin{equation}
\begin{aligned}
\pi_{\psi}(\mathbf{a}_t|s_t)&=\prod_{m=1}^{M}\text{Bernoulli}\!\left(a_{t,m};p_{t,m}\right),\\
p_{t,m}&=\sigma\!\left(\ell_{t,m}/\tau_t\right),
\end{aligned}
\end{equation}
where $\sigma(\cdot)$ is the sigmoid function and $\tau_t$ is a temperature parameter. Sampling from this distribution yields the RU activation vector $\mathbf{a}_t$, which is executed directly in the environment. This formulation maps naturally to RU sleep scheduling: each dimension corresponds to turning one RU active/sleep, and the stochastic policy allows PPO to explore different activation patterns under the same state.

\subsubsection{Exploration Design in RU Scheduling}
To avoid premature saturation of Bernoulli probabilities (i.e., $p_{t,m}\!\rightarrow\!0$ or $1$ leading to near-zero entropy and poor exploration), we incorporate several exploration mechanisms during training. First, we inject Gaussian noise into the actor logits before sampling,
\begin{equation}
    \tilde{\boldsymbol{\ell}}_t = \boldsymbol{\ell}_t + \boldsymbol{\xi}_t,
    \quad \boldsymbol{\xi}_t \sim \mathcal{N}(\mathbf{0}, \sigma_{\ell,t}^2 \mathbf{I}),
\end{equation}
and apply temperature scaling through $\tau_t$ to control stochasticity. Both $\sigma_{\ell,t}$ and $\tau_t$ are annealed over training to transition from exploration to exploitation. Second, an entropy bonus is added to the objective to encourage diverse RU activation decisions at early stages. These mechanisms are particularly important in our scenario because RU control is high-dimensional and strongly coupled via traffic load, user association, and QoS constraints; insufficient exploration may cause the policy to collapse to a fixed RU pattern that fails to adapt to changing UE distributions.

\subsubsection{Rollout Collection and Advantage Estimation with Episode Boundaries}
PPO is trained on on-policy rollouts collected by interacting with the same environment dynamics used by TD3, including mobility, handovers, and the RU-specific energy/carbon models. At each time step, the transition $(s_t, \mathbf{a}_t, r_t, s_{t+1})$ is recorded along with the log-probability $\log \pi_{\psi}(\mathbf{a}_t|s_t)$ and value estimate $V_{\psi}(s_t)$. We collect rollouts for a fixed number of steps and then perform multiple epochs of PPO updates.

To compute low-variance policy gradients, we use generalized advantage estimation (GAE). Importantly, in our implementation, advantage accumulation is reset at episode boundaries (either environment termination or time-limit truncation), preventing leakage of advantages across independent episodes:
\begin{equation}
\begin{aligned}
\hat{A}_t&=\sum_{l=0}^{\infty}(\gamma\lambda)^l\delta_{t+l},\\
\delta_t&=r_t+\gamma V_{\psi}(s_{t+1})-V_{\psi}(s_t).
\end{aligned}
\end{equation}
This is crucial for RU scheduling since episodes are time-limited and the environment may reset due to simulation horizon; improper bootstrapping would distort the estimated returns and destabilize training.

\subsubsection{Clipped PPO Objective for RU Sleep Decisions}
Given a batch of rollout data, PPO updates the policy by maximizing the clipped surrogate objective:
\begin{equation}
\mathcal{L}_{\text{PPO}}(\psi)=\mathbb{E}\Bigl[\min\Bigl(\rho_t(\psi)\hat{A}_t,
\text{clip}\!\left(\rho_t(\psi),1-\epsilon,1+\epsilon\right)\hat{A}_t
\Bigr)\Bigr],
\end{equation}
where
\begin{equation}
\rho_t(\psi)=\exp\Bigl(\log\pi_{\psi}(\mathbf{a}_t|s_t)
-\log\pi_{\psi_{\text{old}}}(\mathbf{a}_t|s_t)\Bigr),
\end{equation}
and $\epsilon$ is the clipping parameter. The value function is trained with a squared error loss, and an entropy regularizer is used to promote exploration:
\begin{equation}
\mathcal{L}(\psi)=-\mathcal{L}_{\text{PPO}}(\psi)
+c_v\mathbb{E}\!\left[(\hat{R}_t-V_{\psi}(s_t))^2\right]
-c_H\mathbb{E}\!\left[\mathcal{H}\!\left(\pi_{\psi}(\cdot|s_t)\right)\right],
\end{equation}
where $\hat{R}_t$ is the estimated return, $\mathcal{H}(\cdot)$ denotes entropy, and $c_v$ and $c_H$ control the value and entropy terms, respectively. In the RU sleep control context, this objective updates the RU activation probabilities so that actions yielding lower energy/carbon cost under QoS constraints become more likely, while the clipped update prevents abrupt policy shifts that could cause unstable RU switching.

\subsubsection{Stability Measures and Implementation Details}
To improve numerical stability in high-dimensional binary control, we clamp logits to bounded ranges before computing log-probabilities and apply gradient norm clipping during optimization. Moreover, reward normalization is applied per rollout to reduce scale sensitivity and to stabilize the advantage estimates. The actor and critic are optimized jointly using separate learning rates for the shared backbone/actor head and the critic head. These design choices help mitigate training collapse (e.g., entropy going to zero) that can occur when the policy quickly becomes overconfident in a particular RU activation pattern.

Overall, PPO observes the same state representation as TD3 and outputs a stochastic RU activation vector $\mathbf{a}_t$ that is executed directly in the environment. This provides a strong and widely adopted policy-gradient baseline for evaluating the suitability of stochastic versus deterministic control paradigms in sustainability-aware RU sleep scheduling.

\section{Simulation Environment}

\begin{figure}[!t]
    \centering

    \subfloat[2 Renewable RUs]{%
        \includegraphics[width=0.32\columnwidth]{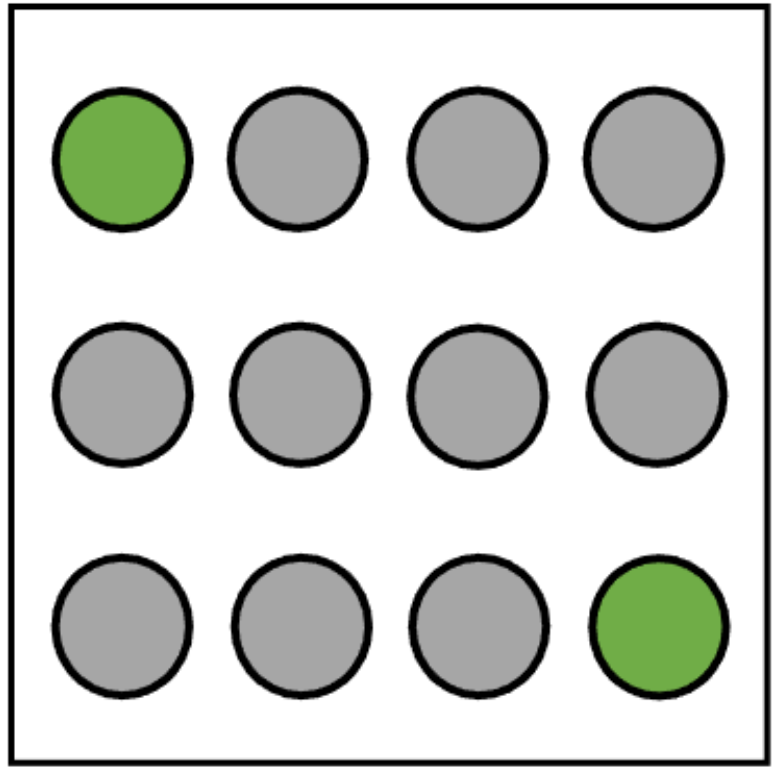}}
    \hfill
    \subfloat[4 Renewable RUs (Corner)]{%
        \includegraphics[width=0.32\columnwidth]{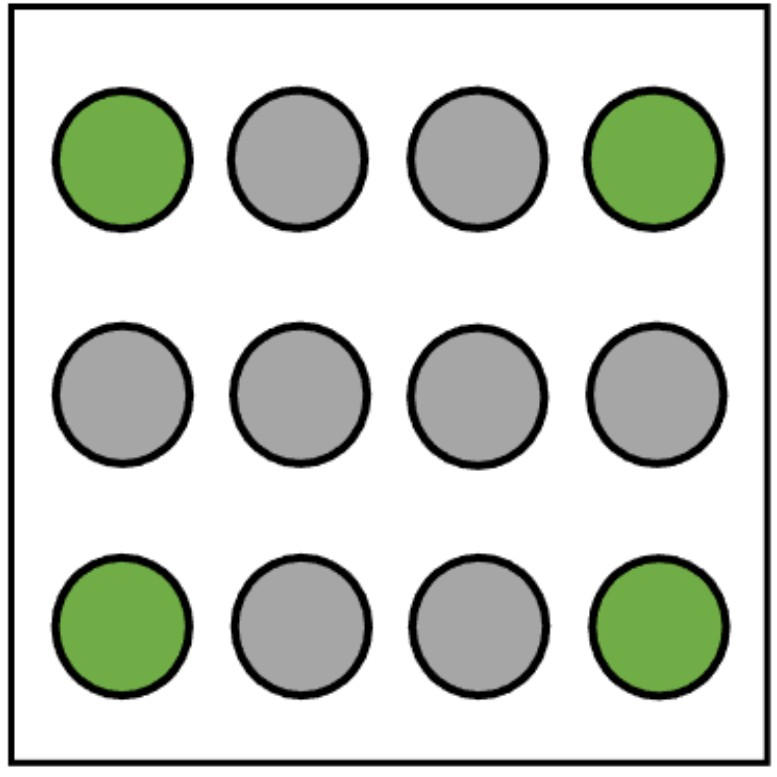}}
    \hfill
    \subfloat[4 Renewable RUs (Middle)]{%
        \includegraphics[width=0.32\columnwidth]{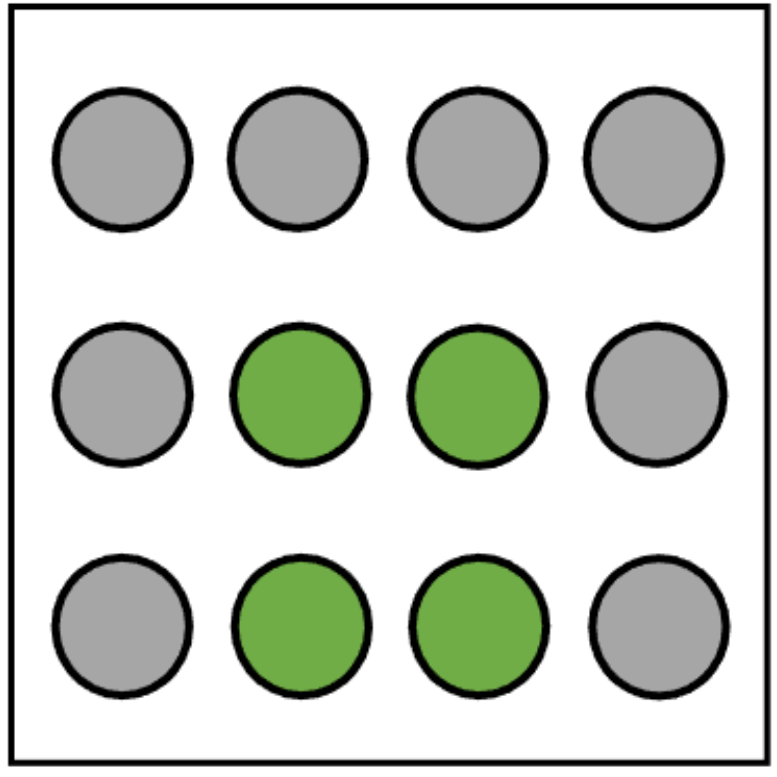}}

    \par\medskip

    \subfloat[6 Renewable RUs (Random)]{%
        \includegraphics[width=0.32\columnwidth]{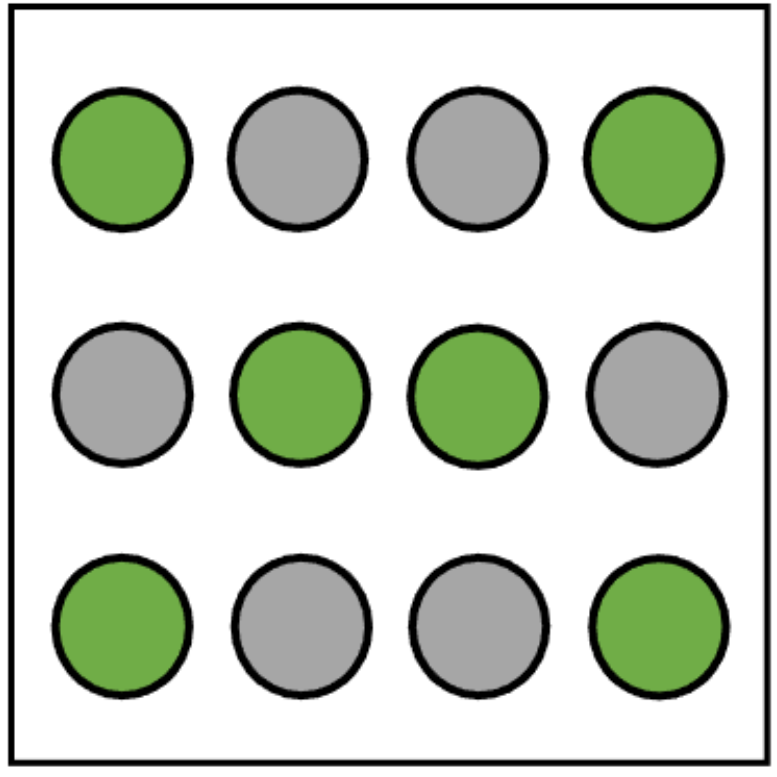}}
    \hfill
    \subfloat[6 Renewable RUs (Even)]{%
        \includegraphics[width=0.32\columnwidth]{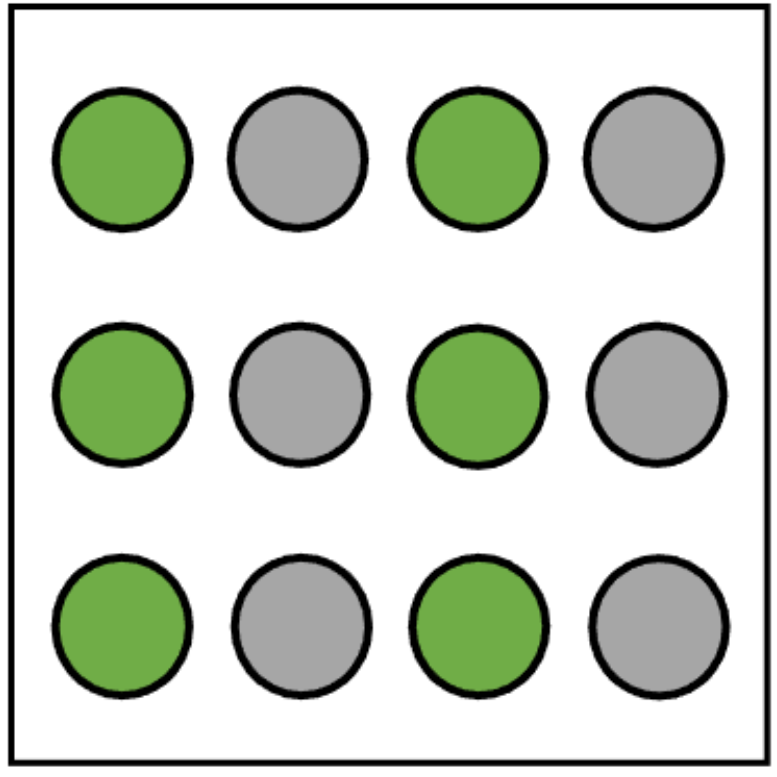}}
    \hfill
    \subfloat[8 Renewable RUs]{%
        \includegraphics[width=0.32\columnwidth]{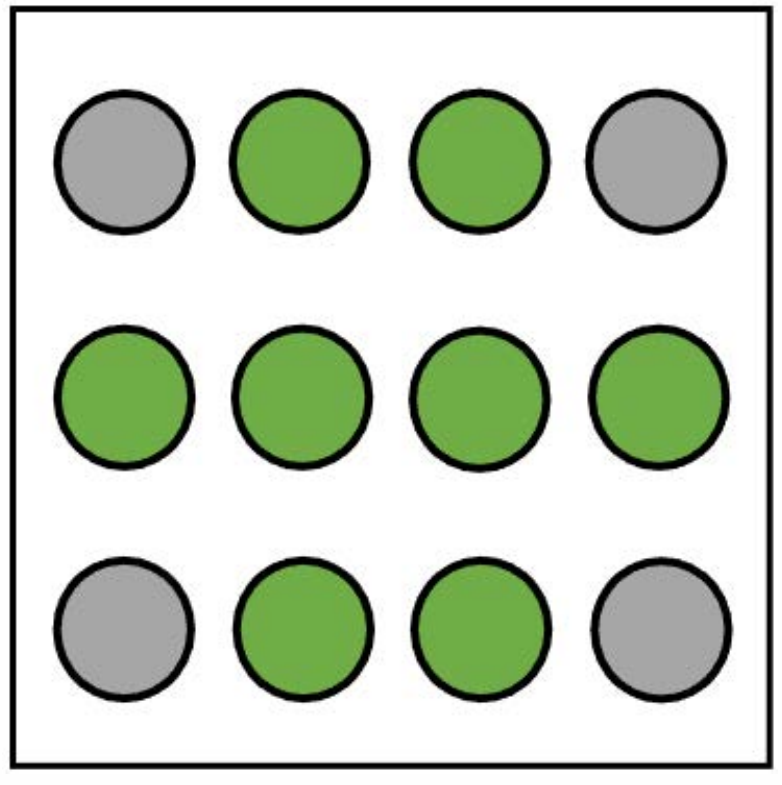}}

    \caption{Spatial deployment scenarios with different renewable RU penetration levels and layouts. Green circles indicate renewable-powered RUs, while gray circles represent grid-powered RUs. These scenarios are used to evaluate the robustness of RU sleep scheduling policies under varying renewable availability and spatial distributions.}
    \label{fig:ru_scenarios}
\end{figure}

\subsection{Simulation Settings}

In our simulations, each scenario is modeled as an \ac{O-RAN} environment where $M$ \acp{RU} serve $K$ \acp{UE} within a square area of size $L \times L$\,m$^2$. The considered scenarios explicitly differentiate between renewable-powered RUs and grid-powered RUs, enabling the evaluation of sustainability-aware RU sleep control under heterogeneous energy sources. Renewable-powered RUs are assumed to operate with zero operational carbon emissions, while grid-powered RUs incur both energy consumption and carbon emissions according to the adopted energy and carbon models.

In practical \ac{O-RAN} deployments, the Near-RT \ac{RIC} is typically hosted at edge cloud or regional edge data centers, with latency budgets ranging from 10\,ms to 1\,s. In our system model, the inference of the DRL agents is executed within the Near-RT RIC as an xApp, ensuring that RU sleep/active decisions satisfy the real-time control requirements. Model training is performed offline using simulated interaction data, while the trained policies are deployed for online inference. This deployment setting aligns with current O-RAN architectural principles and ensures realistic decision-making latency.

As illustrated in Fig.~\ref{fig:ru_scenarios}, multiple spatial layouts are considered to examine the impact of renewable penetration and spatial distribution. The scenarios include different numbers of renewable-powered RUs (e.g., 2, 4, 6, and 8), as well as different spatial arrangements such as centralized, evenly distributed, and random placements. These layouts are designed to evaluate the robustness of the learned policies under varying renewable availability and spatial heterogeneity.

\acp{UE} move according to a periodic mobility model with constant speed
$v = v_{\text{avg}} \pm v_{\text{std}}$, where the mean speed is $v_{\text{avg}} = 2$\,m/s and the standard deviation is $v_{\text{std}} = 0.5$\,m/s. Individual UE speeds are randomly drawn from this range. UEs periodically travel from the boundary of the service area toward the center and back, forming a structured cyclic mobility pattern. This mobility model induces time-varying traffic loads and handovers across RUs, which is critical for evaluating adaptive RU sleep control policies.

The wireless channel model follows the 3GPP \ac{UMi} specification~\cite{3gppTR38901} with both \ac{LOS} and \ac{NLOS} conditions. For LOS links, the path loss is given by
\begin{equation}
\begin{aligned}
PL_{m,k}^{\text{LOS}} =
\begin{cases}
32.4 + 21 \log_{10}(d_{m,k}) \\
\quad {}+ 20 \log_{10}(f), & d_{m,k} \le d_{\text{BP}}, \\[4pt]
32.4 + 40 \log_{10}(d_{m,k}) \\
\quad {}+ 20 \log_{10}(f) - 9.5, & d_{m,k} > d_{\text{BP}} ,
\end{cases}
\end{aligned}
\end{equation}
where $d_{m,k}$ denotes the distance between RU $m$ and UE $k$, $f$ is the carrier frequency in GHz, and $d_{\text{BP}} = \frac{4 h_{\text{RU}} h_{\text{UE}} f}{c}$ is the breakpoint distance. For NLOS links, the path loss is modeled as
\begin{equation}
\begin{aligned}
PL_{m,k}^{\mathrm{NLOS}} =
&\, 35.3 + 22.4 \log_{10}(d_{m,k}) \\
&\, {}+ 21.3 \log_{10}(f)
      - 0.3\bigl(h_{\mathrm{UE}} - 1.5\bigr).
\end{aligned}
\end{equation}
The LOS probability is defined as
\begin{equation}
P_{\text{LOS}} = \min \left( \frac{18}{d_{m,k}}, 1 \right)
\left(1 - e^{-d_{m,k}/36} \right)
+ e^{-d_{m,k}/36},
\end{equation}
with $P_{\text{NLOS}} = 1 - P_{\text{LOS}}$, and $P_{\text{LOS}}=1$ when $d_{m,k}<18$\,m.

\begin{table}[!t]
\centering
\footnotesize
\caption{Simulation Parameters}
\label{tab:sim_params}
\begin{tabularx}{\columnwidth}{@{}>{\raggedright\arraybackslash}X c@{}}
\toprule
\textbf{Parameter} & \textbf{Value} \\ \midrule

\midrule
\multicolumn{2}{c}{\textbf{DRL Training Parameters}} \\
TD3 actor learning rate ($\alpha_{\pi}$) & $1\times10^{-4}$ \\
TD3 critic learning rate ($\alpha_{Q}$) & $1\times10^{-3}$ \\
PPO learning rate ($\alpha_{\text{PPO}}$) & $3\times10^{-4}$ \\
PPO clipping parameter ($\epsilon$) & 0.2 \\
GAE parameter ($\lambda$) & 0.95 \\
Entropy coefficient ($c_H$) & 0.01 \\
Value loss coefficient ($c_V$) & 0.5 \\
Mini-batch size ($B$) & 128 \\
Replay buffer size (TD3 only) & $5\times10^{4}$ \\
Discount factor ($\gamma$) & 0.99 \\
Training episodes ($N_{\text{eps}}$) & 2000 \\
Soft update coefficient (TD3, $\tau$) & 0.01 \\
Random seed & 42 \\

\midrule
\multicolumn{2}{c}{\textbf{Network and Channel Parameters}} \\
Carrier frequency ($f$) & 2 GHz \\
RU height ($h_{\text{RU}}$) & 15 m \\
UE height ($h_{\text{UE}}$) & 1.7 m \\
Network size ($L$) & $1000$ m \\
Number of RUs ($M$) & $12$ \\
Number of UEs ($K$) & $40$, $80$ \\
Noise power density ($\sigma_n^2$) & $-174$ dBm/Hz \\

\midrule
\multicolumn{2}{c}{\textbf{Mobility and Traffic Parameters}} \\
Average UE speed ($v_{\text{avg}}$) & 2 m/s \\
Std. deviation of UE speed ($v_{\text{std}}$) & 0.5 m/s \\
Minimum data rate ($R_{\text{min}}$) & 6 Mbps \\

\midrule
\multicolumn{2}{c}{\textbf{Energy and Carbon Model Parameters}} \\
Active mode RU power ($P^{\text{active}}$) & 20 W \\
Sleep mode RU power ($P^{\text{sleep}}$) & 5 W \\
Mode transition power ($P^{\text{trans}}$) & 3 W \\
Maximum transmit power ($P_{\text{TX}}$) & 1 W (30 dBm) \\
Power amplifier efficiency ($\eta$) & 0.5 \\
Reward weights $(w_1,w_2,w_3)$ & $(1, 3, 10)$ \\
Reward weights $(w_4,w_5)$ & $(1, 2.5)$ \\

\bottomrule
\end{tabularx}
\end{table}

Each training episode consists of 200 time steps. To account for RU state transition latency and to match practical network operation timescales, each time step corresponds to 1.75\,s. This value is motivated by the slow-timescale BS sleep-control model in~\cite{ju2022energy}, where the decision interval is selected according to the coherence time of large-scale fading and BS transition latency. In the present work, 1.75~s is used as the interval between two consecutive RU activation decisions and provides time for telemetry collection, policy inference,
E2 signaling, and RU actuation. RU sleep/active decisions, energy consumption measurement, and carbon emission accounting are all performed at this temporal granularity. The instantaneous energy and carbon emission values are recorded at each step, while episode-level metrics are obtained by accumulation over the entire episode.

We evaluate two representative \ac{DRL} approaches for centralized RU sleep control: the deterministic policy-based \ac{TD3} and the stochastic policy-gradient-based \ac{PPO}. Both agents share the same state representation, action space, and environment dynamics to ensure a fair and consistent comparison. 
To investigate the impact of sustainability-aware optimization, three different reward formulations are considered, are summarized in Table \ref{tab:reward_design}. AG1 jointly accounts for normalized energy consumption, carbon emissions, and QoS satisfaction, enabling a balanced trade-off between efficiency and sustainability. AG2 focuses on carbon-aware optimization by considering carbon emissions and QoS satisfaction only, while AG3 represents a conventional energy-driven design that optimizes energy consumption and QoS without explicitly incorporating carbon emissions \cite{liang2026scalable}. This setup allows us to systematically analyze the interaction and trade-offs between energy efficiency and carbon reduction.Detailed network architectures, optimizer configurations and system parameters are summarized in Table~\ref{tab:sim_params}.

To provide a non-learning baseline, a green-aware hysteresis heuristic is implemented. The heuristic follows five rules. First, any RU serving at least one UE is kept active. Second, an RU is considered idle if it has no associated UEs and its PRB utilisation is below the idle threshold \(U_{\mathrm{off}}\). Third, an idle RU is switched off after it has remained idle for \(T_{\mathrm{off}}\) consecutive time steps. Fourth, if the average user satisfaction falls below \(S_{\min}\), or if any active RU has PRB utilisation above the congestion threshold \(U_{\mathrm{on}}\), the heuristic activates nearby sleeping RUs to relieve congestion. When multiple candidate RUs are available, renewable-powered RUs are prioritised before grid-powered RUs. Finally, at least one RU is always kept active as a safety condition.

The idle threshold is set to \(U_{\mathrm{off}}=0.20\), the congestion threshold is set to \(U_{\mathrm{on}}=0.90\), and the minimum average user satisfaction is set to \(S_{\min}=0.95\). The hysteresis duration is \(T_{\mathrm{off}}=0\), meaning that an idle RU can be switched off immediately once the idle condition is satisfied. To avoid activating too many RUs in one control step, the maximum number of additionally activated RUs is limited to two. The heuristic is evaluated over 50 episodes, with 200 time steps per episode.

\begin{table}[!t]
\centering
\footnotesize
\caption{Comparison of Reward Function Designs}
\label{tab:reward_design}
\begin{tabular}{@{}cl@{}}
\toprule
\textbf{Reward function} & \textbf{Description} \\ \midrule
AG1 & Energy + Carbon + QoS \\
AG2 & Carbon + QoS  \\
AG3 & Energy + QoS  \\
\bottomrule
\end{tabular}
\end{table}

\subsection{Numerical Results}

\begin{figure}[!t]
    \centering

    \subfloat[40 UEs, Ag 1]{%
        \includegraphics[width=0.48\columnwidth]{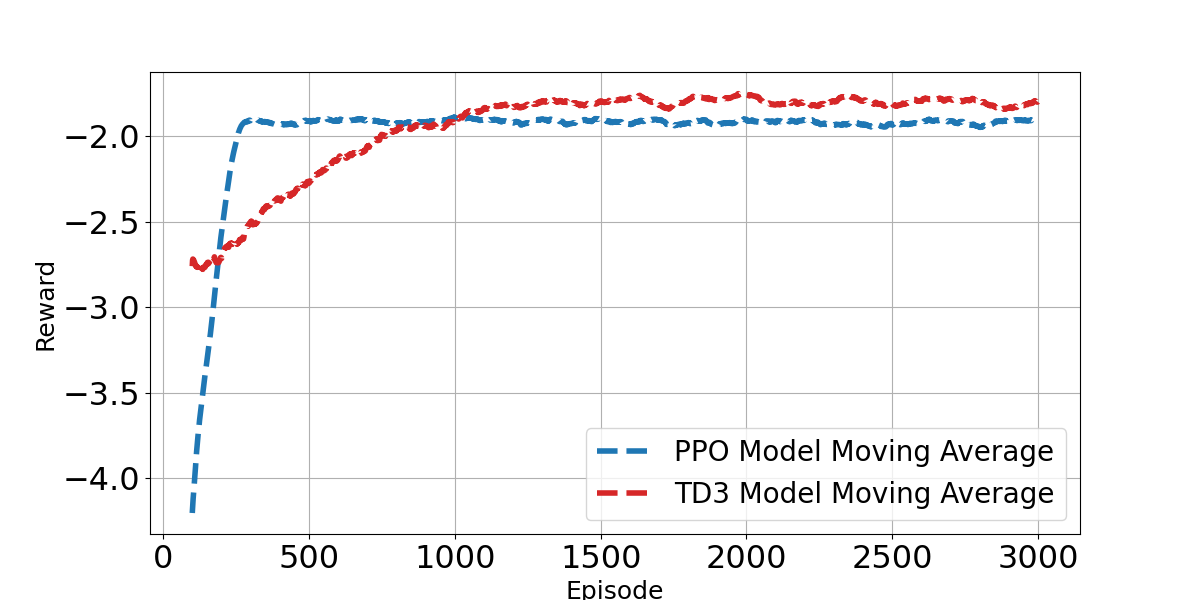}}
    \hfill
    \subfloat[40 UEs, Ag 2]{%
        \includegraphics[width=0.48\columnwidth]{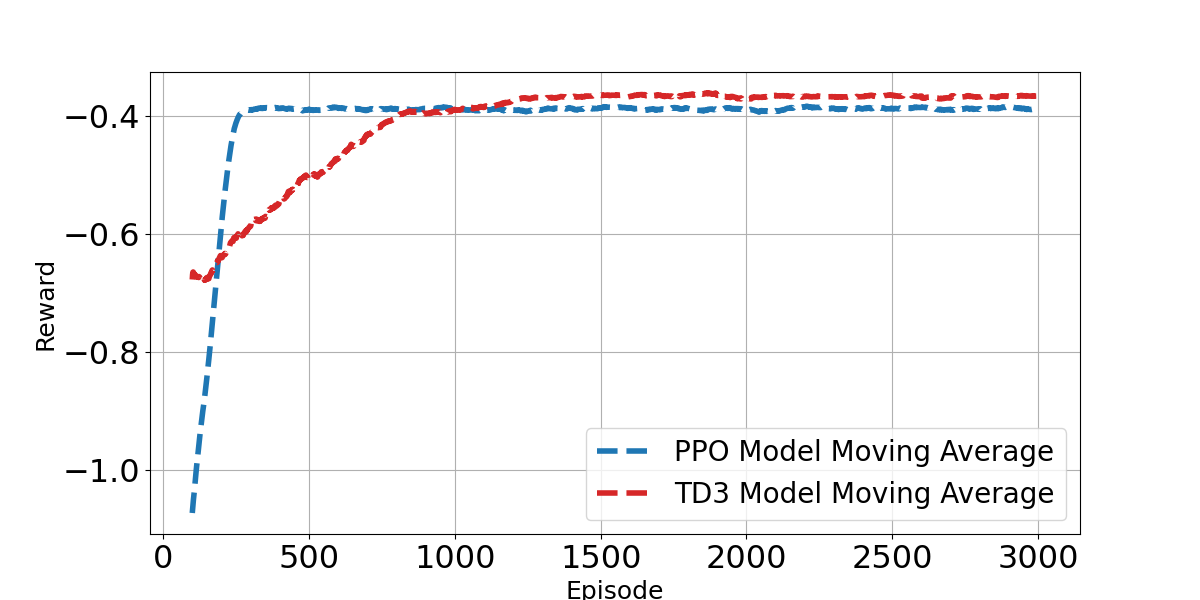}}

    \par\medskip

    \subfloat[80 UEs, Ag 1]{%
        \includegraphics[width=0.48\columnwidth]{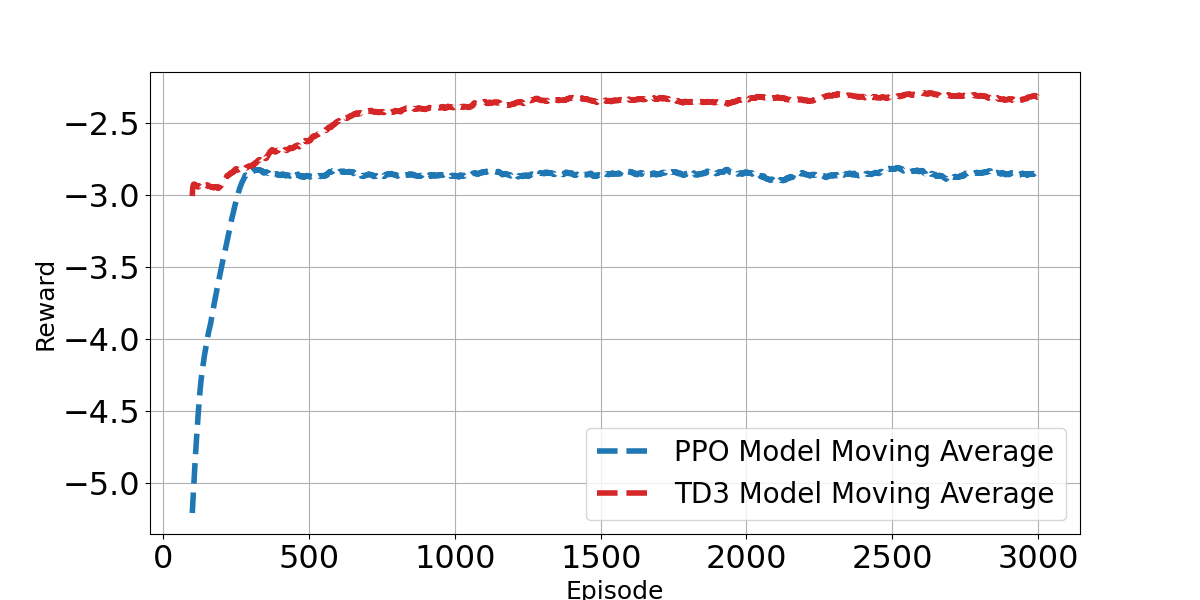}}
    \hfill
    \subfloat[80 UEs, Ag 2]{%
        \includegraphics[width=0.48\columnwidth]{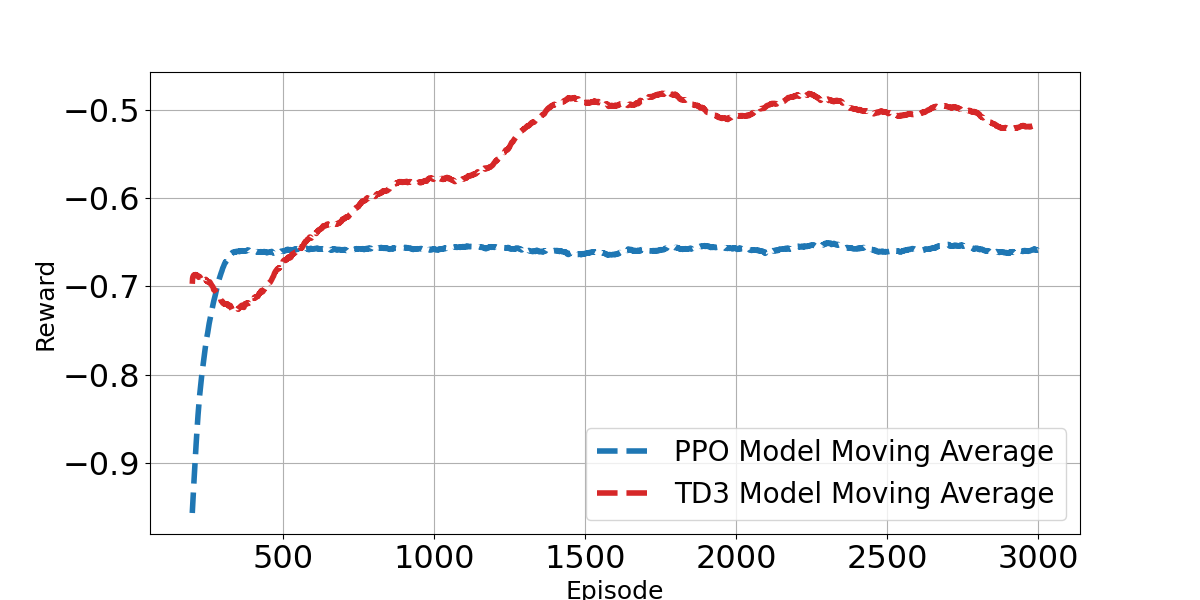}}

    \caption{Training convergence comparison between PPO and TD3 under the 6 renewable RU (random layout) scenario. The moving-average episode rewards are shown for 40 UE and 80 UE traffic loads, each evaluated with two independent reward functions (AG1 and AG2).}
    \label{fig:train_reward}
\end{figure}

Fig.~\ref{fig:train_reward} compares the training convergence of TD3 and PPO under the 6-renewable-RU random deployment scenario for both 40-UE and 80-UE traffic loads. Two reward formulations, AG1 and AG2, are evaluated. For the moderate traffic case with 40 UEs, both algorithms converge within approximately the first 1000 episodes. PPO shows faster improvement during the initial training phase, which can be attributed to its stochastic exploration over binary RU activation patterns. However, TD3 improves more steadily and reaches higher final rewards under both reward functions. This suggests that, after sufficient exploration, the deterministic off-policy structure of TD3 is more effective in refining RU activation decisions.

Under the heavier 80-UE load, the performance gap between TD3 and PPO becomes more pronounced. The increased traffic demand tightens QoS constraints and strengthens the coupling among RU activation, user association, and PRB utilization. In this more complex setting, PPO converges quickly but tends to stabilize in a suboptimal region, whereas TD3 continues to improve over a longer training horizon and achieves higher asymptotic rewards. This demonstrates the advantage of experience replay and twin-critic learning in handling high-dimensional RU sleep control under stringent service requirements.

The reward formulation also affects the convergence behavior. AG1, which jointly considers energy, carbon emissions, and QoS, generally leads to smoother learning because the energy term regularizes excessive RU activation. In contrast, AG2 focuses more directly on carbon reduction, creating a sharper and more asymmetric reward landscape between renewable- and grid-powered RUs. This makes learning more challenging, especially for PPO. TD3 remains comparatively stable across both reward functions, showing stronger robustness to reward shaping and traffic-load variations.

\begin{figure}[!t]
\centering

\subfloat[2 Renewable RUs]{%
    \includegraphics[width=0.5\columnwidth]{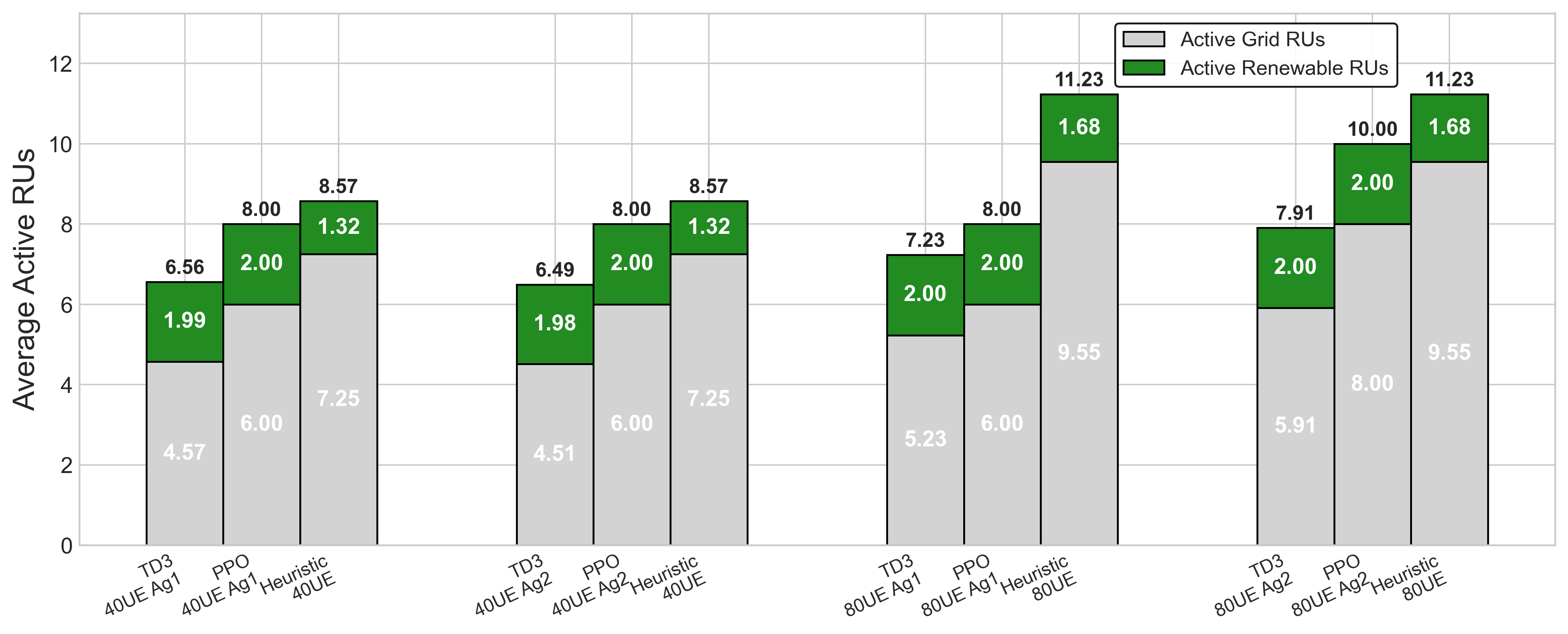}}
\hfill
\subfloat[4 Renewable RUs (Corner)]{%
    \includegraphics[width=0.5\columnwidth]{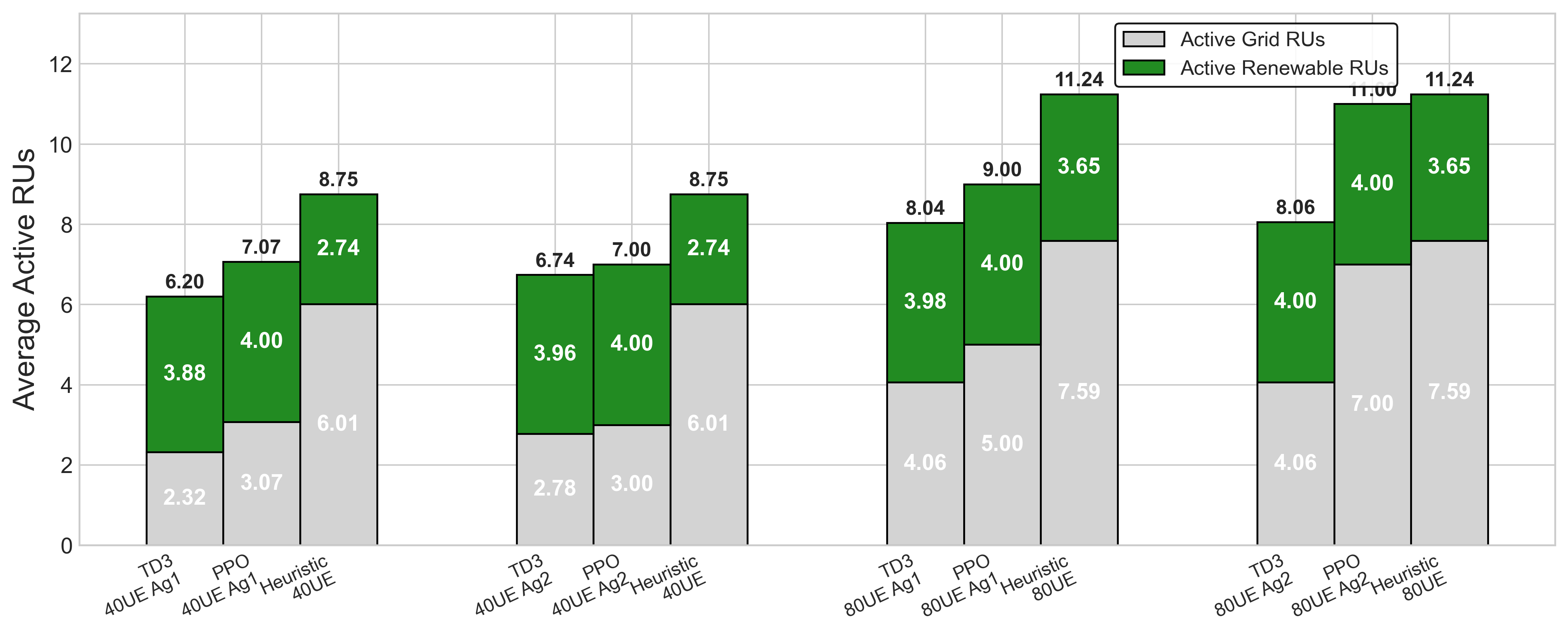}}
    
\par\medskip

\subfloat[4 Renewable RUs (Middle)]{%
    \includegraphics[width=0.5\columnwidth]{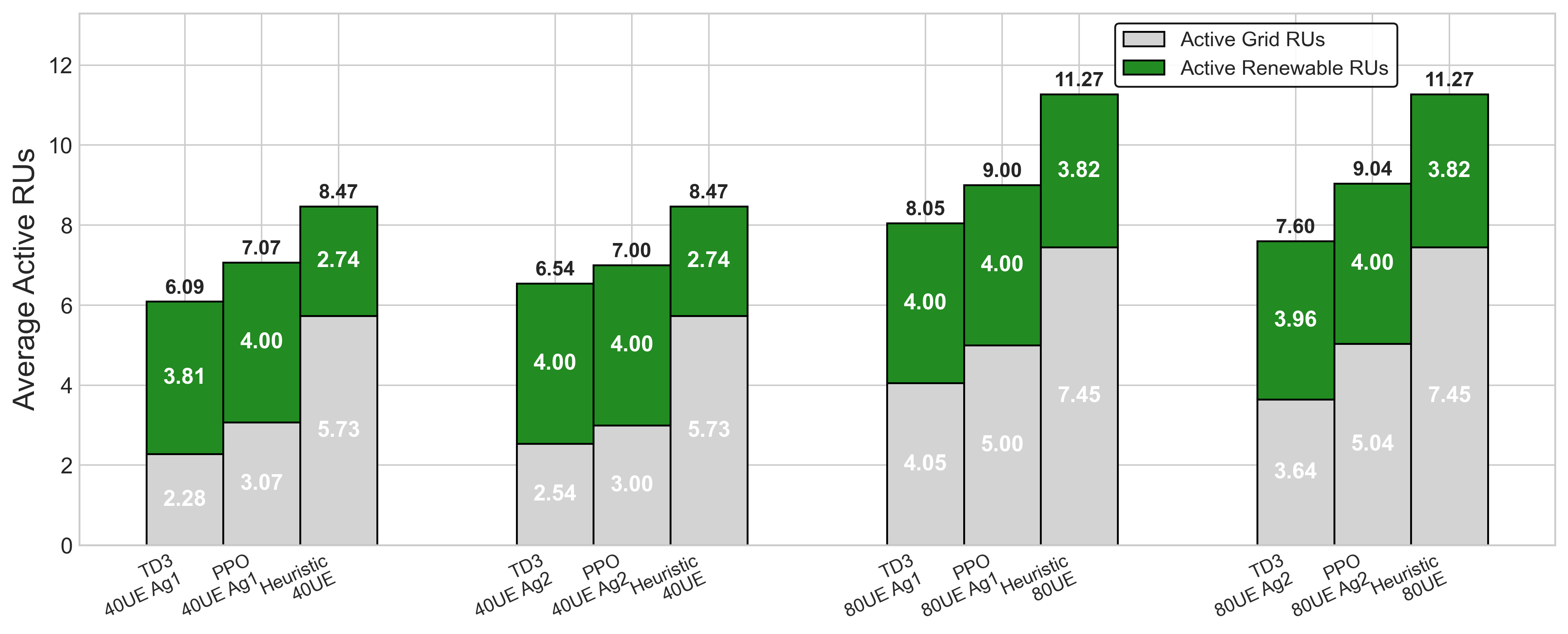}}
\hfill
\subfloat[6 Renewable RUs (Random)]{%
    \includegraphics[width=0.5\columnwidth]{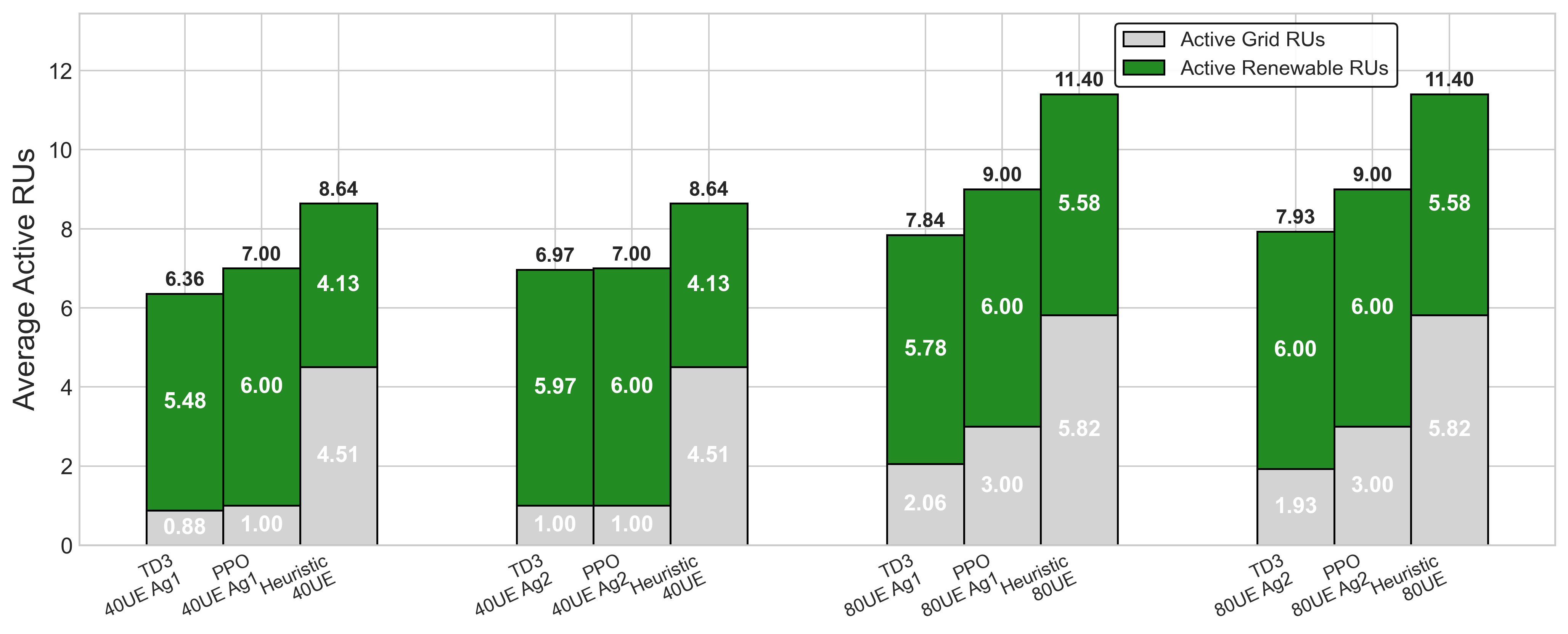}}
    
\par\medskip

\subfloat[6 Renewable RUs (Even)]{%
    \includegraphics[width=0.5\columnwidth]{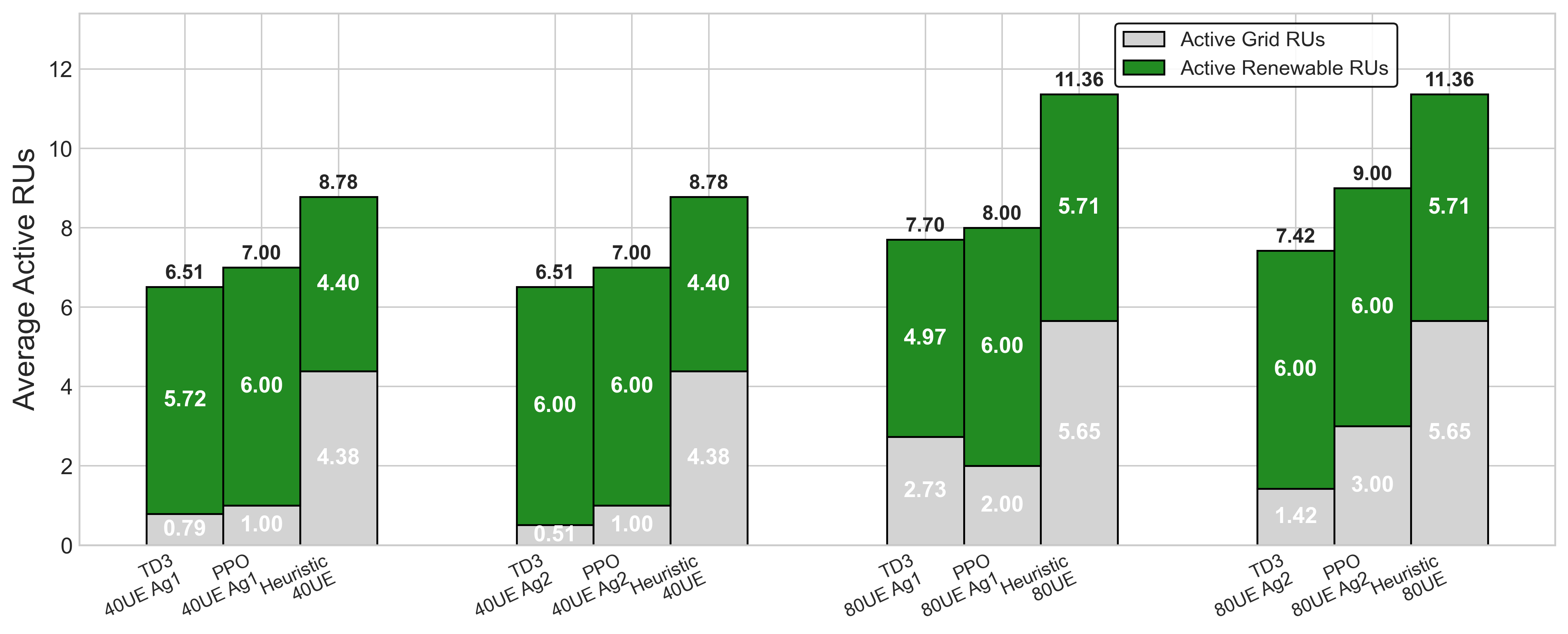}}
\hfill
\subfloat[8 Renewable RUs]{%
    \includegraphics[width=0.5\columnwidth]{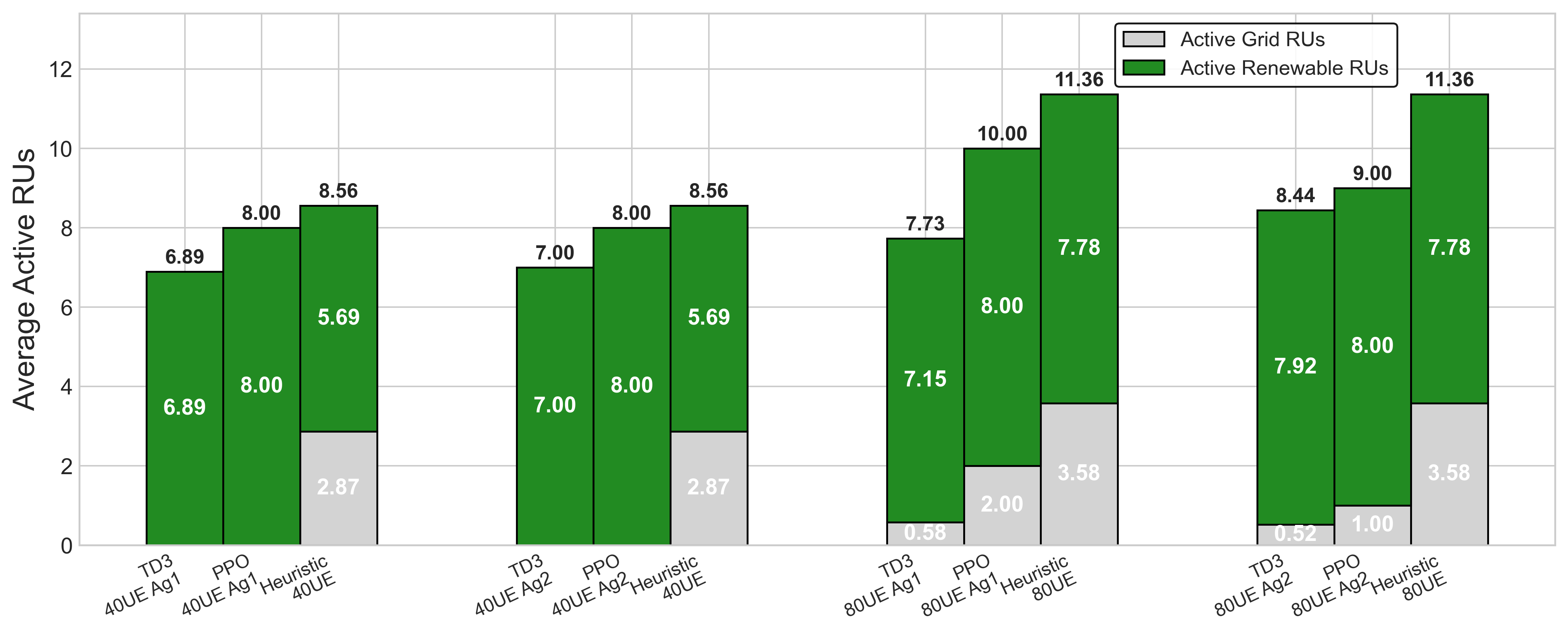}}

\caption{Average number of active renewable-powered and grid-powered RUs under different control strategies for various renewable penetration levels and spatial layouts. Each subfigure corresponds to a different deployment scenario. Results are shown for both 40 UEs and 80 UEs.}
\label{fig:active_ru_summary}
\end{figure}

Fig.~\ref{fig:active_ru_summary} presents the average number of active renewable- and grid-powered RUs under different renewable deployment layouts. A clear operational transition is observed as the number of renewable RUs increases. In low-renewable scenarios, such as the 2-renewable-RU case, learning-based strategies activate nearly all renewable RUs but still rely on grid-powered RUs to satisfy coverage and QoS requirements. As renewable penetration increases, both TD3 and PPO progressively reduce grid-powered activation and shift toward renewable-dominant operation.

Among the evaluated methods, TD3 achieves the most compact activation structure. Across all renewable layouts and traffic loads, TD3 activates on average about one fewer RU than PPO and about 2.5 fewer RUs than the heuristic baseline. The difference is particularly visible under the 80-UE case, where PPO tends to keep more grid-powered RUs active to preserve QoS stability. By contrast, TD3 learns a more selective activation policy that better exploits the spatial distribution of renewable-powered RUs. In the 8-renewable-RU scenario under 40 UEs, TD3 nearly eliminates grid-powered RU activation, indicating strong carbon-aware prioritization.

The influence of reward design is also evident from the activation patterns. AG2 encourages stronger renewable utilization than AG1 because carbon emissions are directly penalized while total energy consumption is not explicitly included. As a result, AG2 may activate additional renewable-powered RUs to reduce grid dependence. AG1, on the other hand, balances energy and carbon objectives, leading to a more compact active set. Compared with the heuristic baseline, both DRL methods are more sensitive to renewable placement and traffic distribution, while TD3 provides the most effective carbon-aware reconfiguration.

\begin{figure}[!t]
    \centering
    \includegraphics[width=1.0\columnwidth]{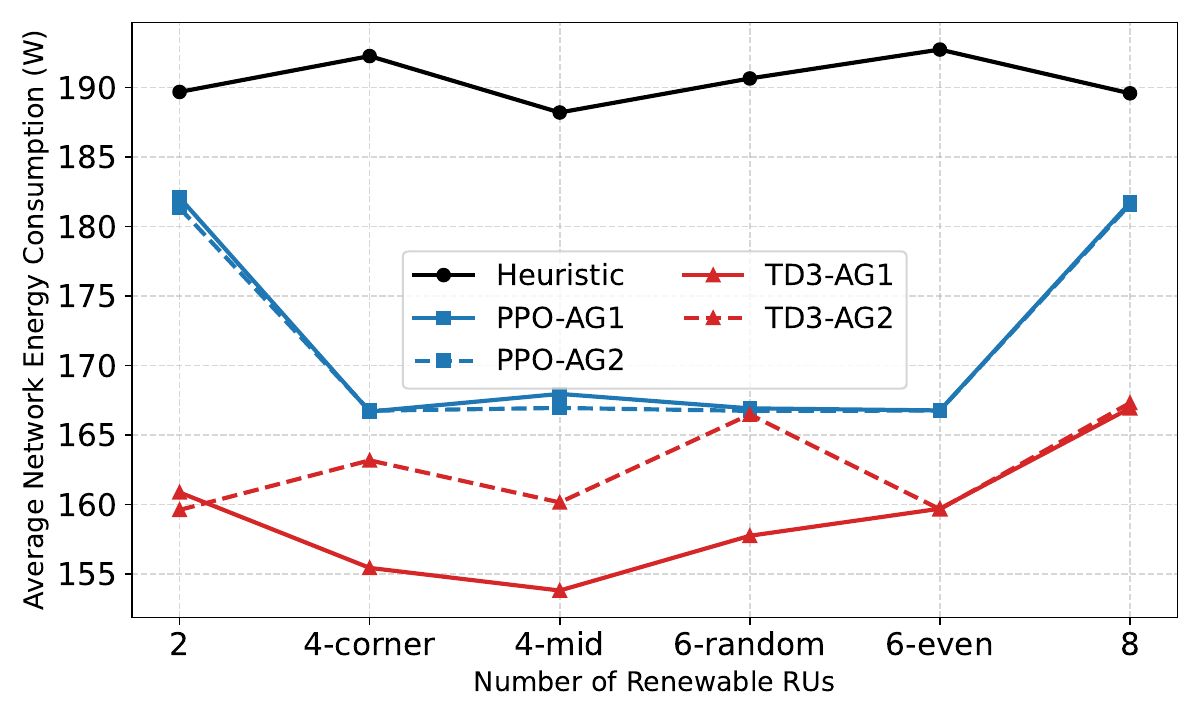}
    \par\medskip
    \includegraphics[width=1.0\columnwidth]{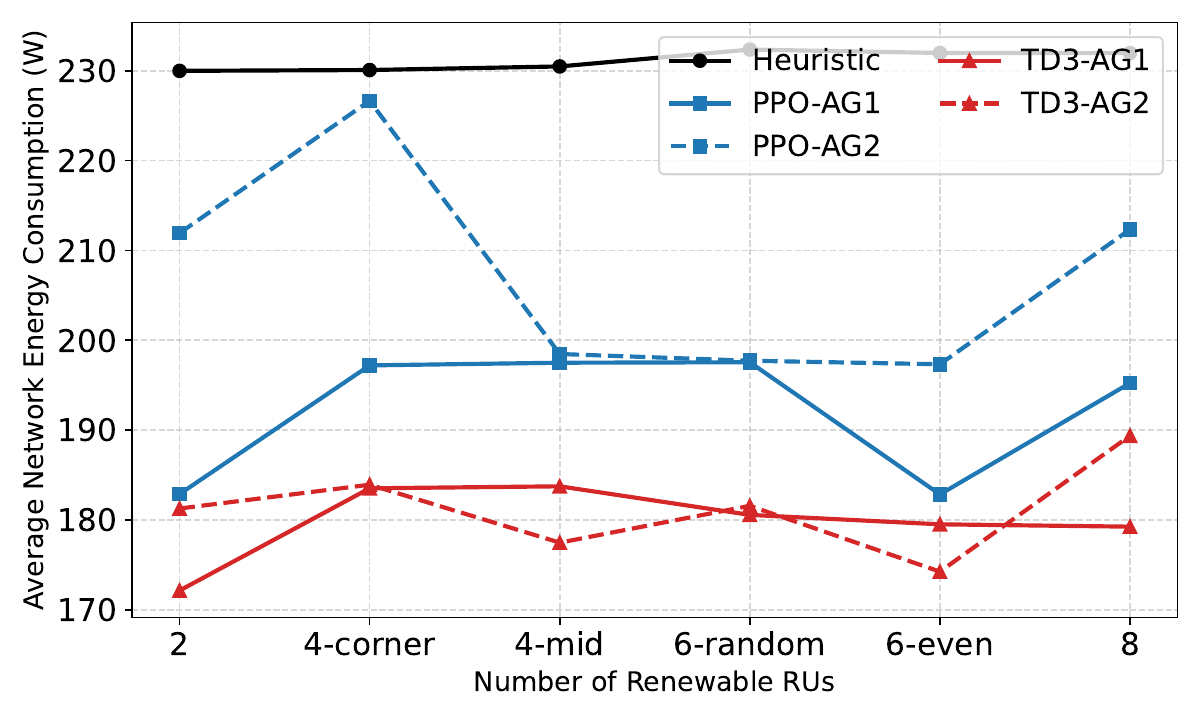}
    \caption{Average network energy consumption under different control strategies as a function of the number of renewable RUs for the scenario with 40 (left) and 80 (right) UEs. Results are shown for TD3, PPO with two reward functions (AG1 and AG2) and heuristic baselines.}
    \label{fig:energy_40ue}
\end{figure}

Fig.~\ref{fig:energy_40ue} presents the average network energy consumption for AG1 and AG2. For 40 UEs, the heuristic baseline has the highest energy consumption, remaining above 180~W in low-renewable scenarios. DRL-based methods reduce energy consumption across all renewable configurations, with TD3 consistently outperforming PPO. In the 2-renewable-RU case, TD3 saves approximately 5--10~W compared with PPO, corresponding to about 3\%--6\% improvement. Similar trends are observed for 80 UEs, where the absolute consumption is higher due to increased traffic demand, but TD3 still achieves typical savings of 5--8~W over PPO.

\begin{figure}[!t]
    \centering
    \includegraphics[width=1.0\columnwidth]{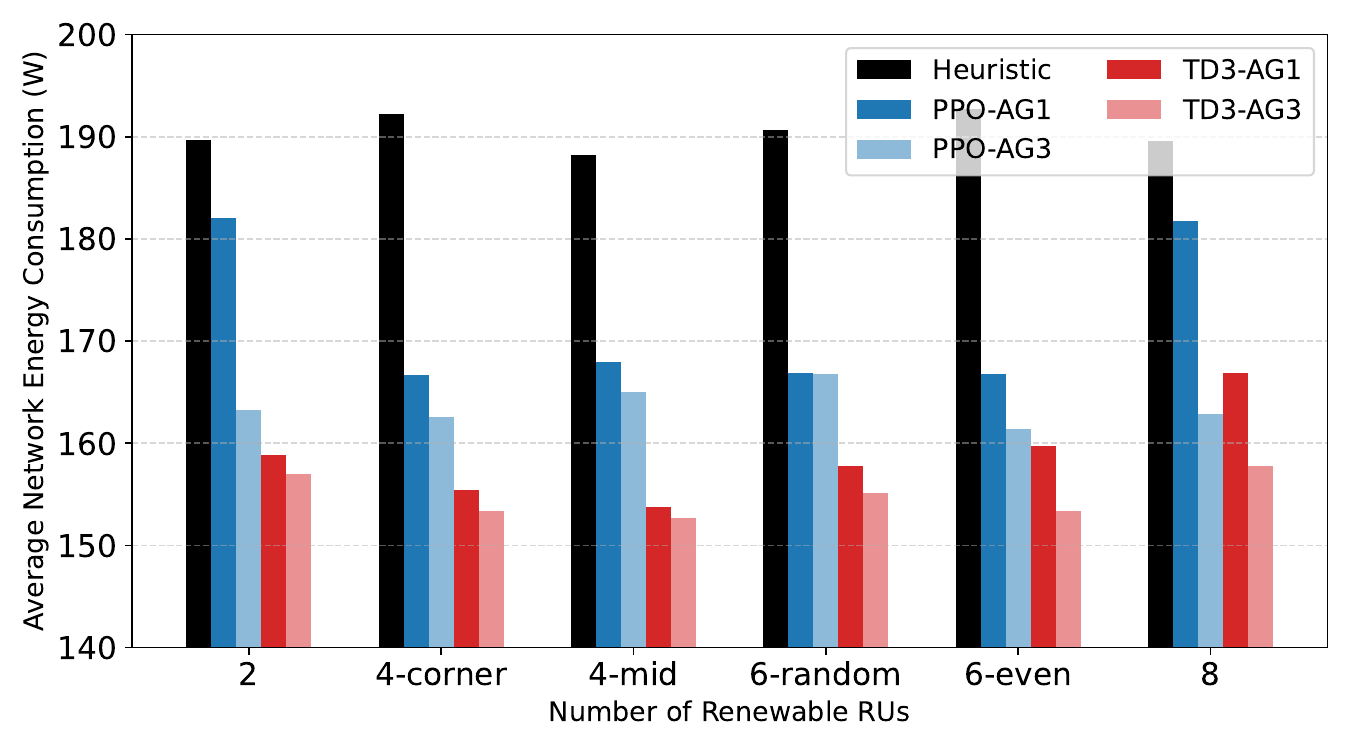}
    \par\medskip
    \includegraphics[width=1.0\columnwidth]{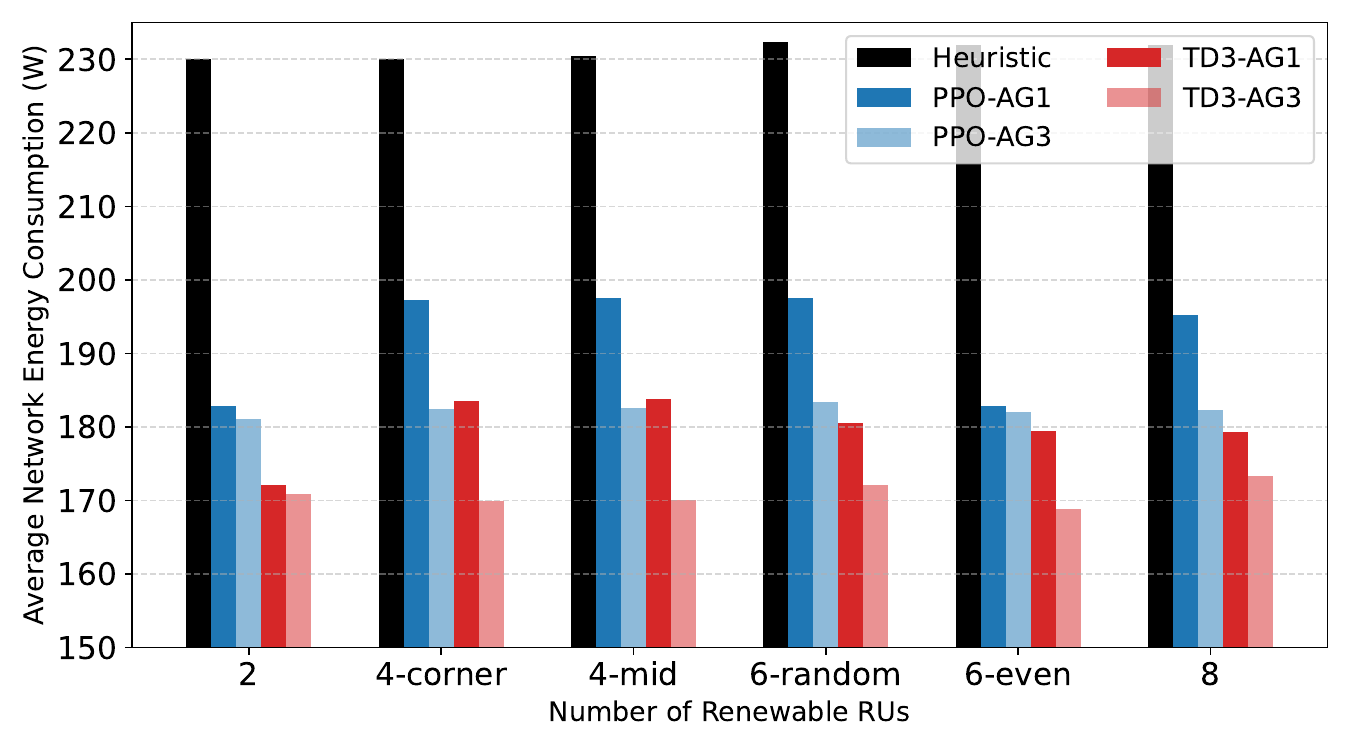}
    \caption{Average network energy consumption under different control strategies as a function of the number of renewable RUs for the scenario with 40 (left) and 80 (right) UEs. Results are shown for TD3, PPO with two reward functions (AG1 and AG3) and heuristic baselines.}
    \label{fig:energy_40ue_bar}
\end{figure}

Reward design also affects energy consumption. AG1 usually consumes less energy than AG2 because it explicitly penalizes total power. AG2, by focusing more strongly on carbon reduction, may activate additional renewable-powered RUs to reduce emissions, leading to slightly higher total energy use. This difference is most visible in moderate-renewable scenarios, where AG1 provides an additional 2--4~W energy reduction.

Fig.~\ref{fig:energy_40ue_bar} compares AG1 with AG3, where AG3 includes energy and QoS terms only. AG3 achieves lower energy consumption than AG1 in most scenarios for both traffic loads, as expected, since it does not consider carbon emissions. However, this lower energy use comes at the cost of higher reliance on grid-powered RUs, as shown later in the carbon-emission results. Therefore, AG3 represents a pure energy-minimization policy, while AG1 provides a more sustainable balance between energy efficiency and carbon reduction.

\begin{figure}[!t]
    \centering
    \includegraphics[width=1.0\columnwidth]{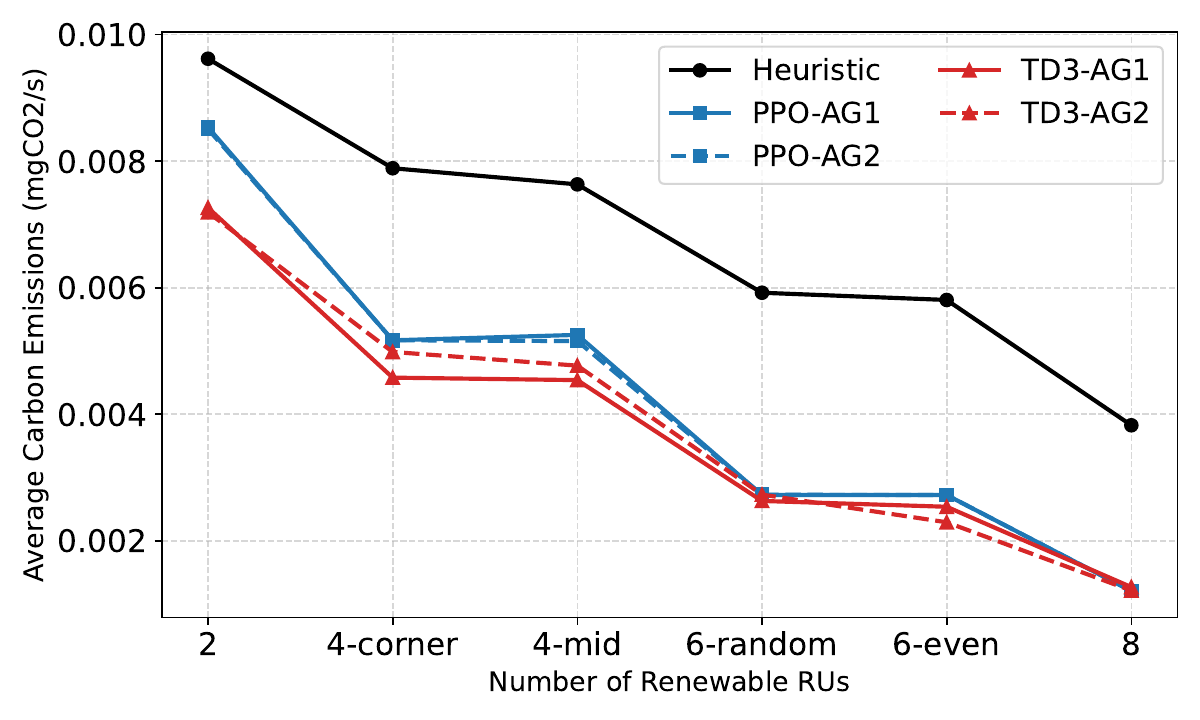}
    \hfill
    \includegraphics[width=1.0\columnwidth]{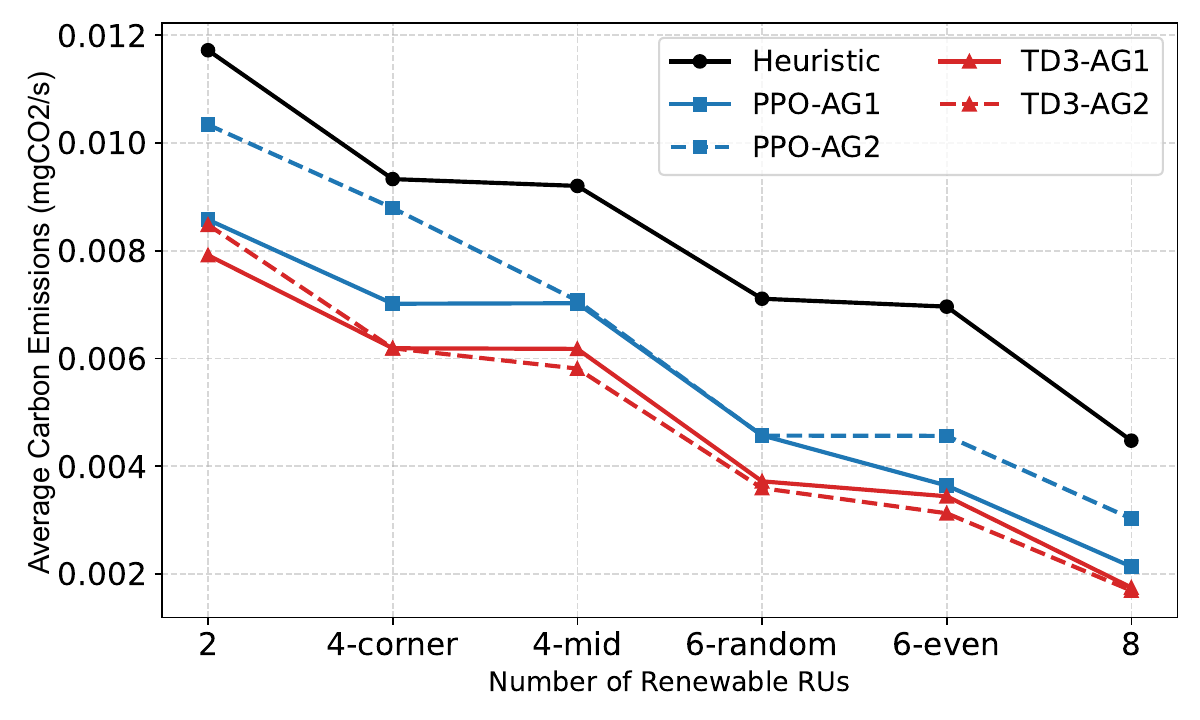}
    \caption{{Average operational carbon emissions under different control strategies as a function of the number of renewable RUs for the scenario with 40 (left) and 80 (right) UEs. The results illustrating how carbon-aware control policies increasingly rely on renewable-powered RUs and reduce grid-dependent emissions as renewable availability grows.}}
    \label{fig:carbon_40ue}
\end{figure}

Fig.~\ref{fig:carbon_40ue} reports the average operational carbon emissions under AG1 and AG2. Carbon emissions decrease monotonically as the number of renewable RUs increases, confirming that renewable penetration shifts the network toward low-carbon operation. For 40 UEs, TD3 achieves the lowest emissions across all renewable configurations. In the 2-renewable-RU case, TD3-AG2 reduces emissions by approximately 8\%--12\% compared with PPO-AG2 and by more than 15\% compared with the heuristic baseline. With higher renewable penetration, TD3 maintains a 6\%--10\% reduction over PPO and over 18\% compared with the heuristic strategy.

For 80 UEs, emissions increase due to heavier traffic, but the relative ranking remains unchanged. TD3-based policies again achieve the lowest emissions, particularly under 4--6 renewable RUs, where TD3-AG2 obtains about 10\% lower emissions than PPO-AG2 and up to 20\% lower emissions than the heuristic baseline. AG2 consistently produces lower emissions than AG1 because it directly prioritizes carbon reduction, while AG1 retains energy awareness and therefore offers a more balanced policy.

\begin{figure}[!t]
    \centering
    \includegraphics[width=1.0\columnwidth]{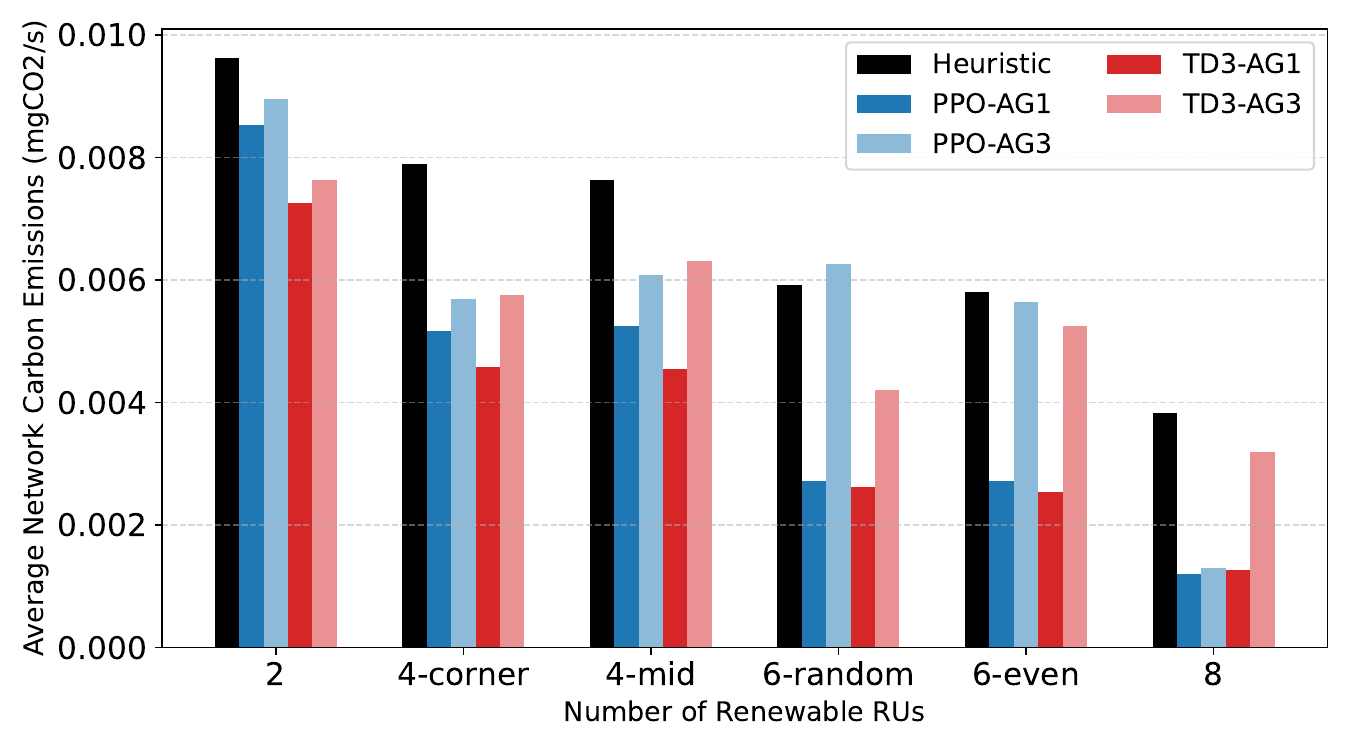}
    \hfill
    \includegraphics[width=1.0\columnwidth]{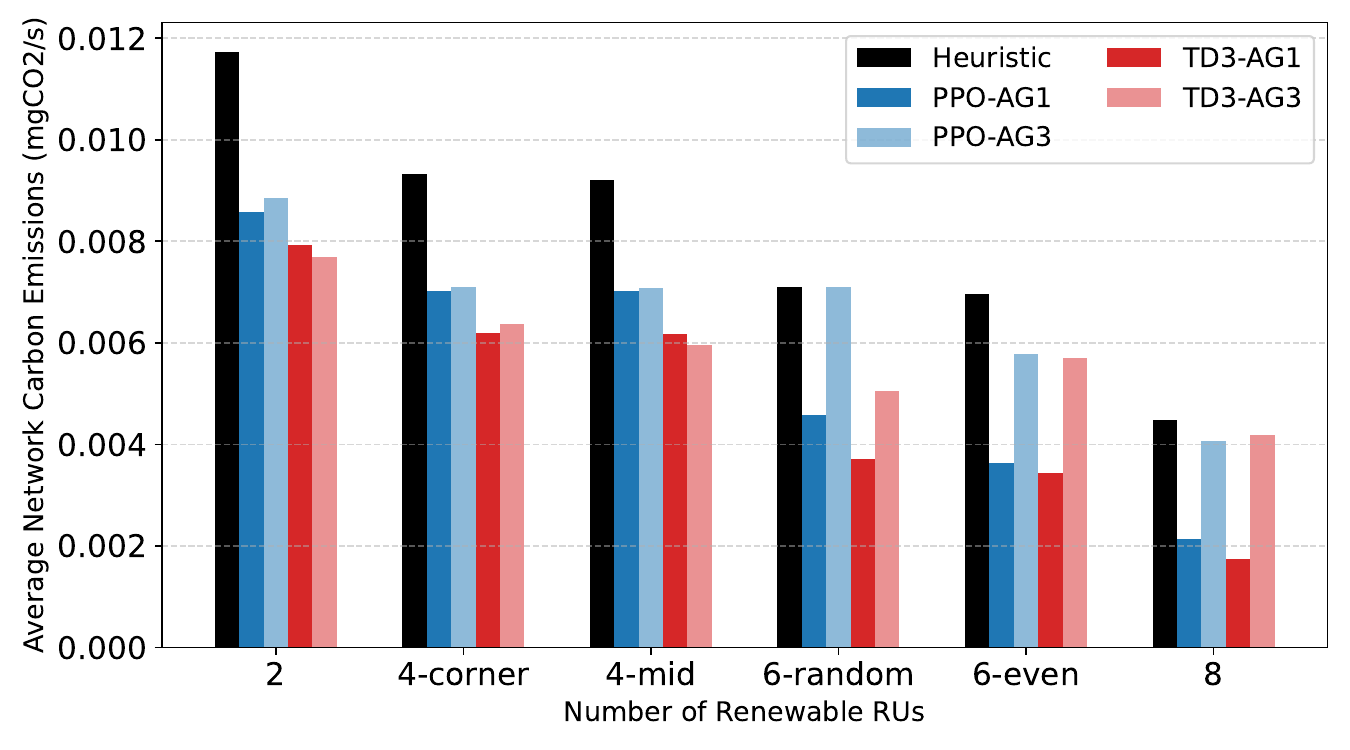}
    \caption{Average operational carbon emissions under different control strategies (AG1 and AG3) as a function of the number of renewable RUs for the scenario with 40 (left) and 80 (right) UEs.}
    \label{fig:carbon_40ue_bar}
\end{figure}

Fig.~\ref{fig:carbon_40ue_bar} compares AG1 and AG3 in terms of carbon emissions. AG1 consistently outperforms AG3 across all renewable deployments and traffic loads, reducing emissions by approximately 12\%--28\%, with an average reduction of about 18\%. This improvement comes from explicitly including carbon emissions in the reward, which encourages the agent to prioritize renewable-powered RUs and reduce grid-powered activation. By contrast, AG3 minimizes energy and QoS penalties only, and therefore may select energy-efficient but carbon-intensive RU configurations.

Together, Figs.~\ref{fig:energy_40ue_bar} and~\ref{fig:carbon_40ue_bar} demonstrate the trade-off between pure energy minimization and carbon-aware control. AG3 provides the lowest energy consumption, whereas AG1 achieves substantial carbon reduction with only a marginal energy increase. These results confirm that explicit carbon-aware reward design is necessary for sustainable RU sleep control in heterogeneous-energy O-RAN systems.

\begin{table*}[!t]
\centering
\caption{{QoS satisfaction achieved by TD3 and PPO under different renewable-RU deployment scenarios and traffic loads. }}
\label{tab:qos_satisfaction}
\footnotesize
\setlength{\tabcolsep}{3.5pt}
\renewcommand{\arraystretch}{1.10}
\begin{tabular}{llcccc}
\toprule
& & \multicolumn{2}{c}{40 UEs}
& \multicolumn{2}{c}{80 UEs} \\
\cmidrule(lr){3-4}\cmidrule(lr){5-6}
Scenario & Reward
& TD3 & PPO
& TD3 & PPO \\
\midrule

2 renewable RUs
& AG1
& $39.70\pm0.51$ (99.25\%)
& $39.93\pm0.13$ (99.83\%)
& $78.60\pm1.21$ (98.25\%)
& $77.88\pm1.33$ (97.35\%) \\

& AG2
& $39.59\pm0.59$ (98.98\%)
& $39.76\pm0.25$ (99.40\%)
& $79.56\pm0.42$ (99.45\%)
& $79.11\pm0.86$ (98.89\%) \\

& AG3
& $39.90\pm0.21$ (99.75\%)
& $39.45\pm0.56$ (98.63\%)
& $79.73\pm0.34$ (99.66\%)
& $79.69\pm0.41$ (99.61\%) \\
\addlinespace

4 renewable RUs (corner)
& AG1
& $39.18\pm0.89$ (97.95\%)
& $39.88\pm0.29$ (99.70\%)
& $78.42\pm1.35$ (98.03\%)
& $79.87\pm0.18$ (99.84\%) \\

& AG2
& $39.80\pm0.56$ (99.50\%)
& $39.90\pm0.26$ (99.75\%)
& $79.20\pm0.74$ (99.00\%)
& $79.78\pm0.26$ (99.72\%) \\

& AG3
& $39.89\pm0.20$ (99.72\%)
& $39.77\pm0.27$ (99.43\%)
& $79.25\pm0.71$ (99.06\%)
& $78.93\pm0.89$ (98.66\%) \\
\addlinespace

4 renewable RUs (middle)
& AG1
& $38.67\pm0.86$ (96.67\%)
& $38.86\pm0.92$ (97.15\%)
& $78.89\pm1.06$ (98.61\%)
& $78.06\pm1.04$ (97.58\%) \\

& AG2
& $39.70\pm0.52$ (99.25\%)
& $38.96\pm0.78$ (97.40\%)
& $78.70\pm1.11$ (98.38\%)
& $76.01\pm1.32$ (95.01\%) \\

& AG3
& $39.77\pm0.33$ (99.43\%)
& $39.45\pm0.56$ (98.63\%)
& $78.97\pm0.78$ (98.71\%)
& $78.93\pm0.89$ (98.66\%) \\
\addlinespace

6 renewable RUs (random)
& AG1
& $39.72\pm0.36$ (99.30\%)
& $39.70\pm0.46$ (99.25\%)
& $79.24\pm0.63$ (99.05\%)
& $79.37\pm0.68$ (99.21\%) \\

& AG2
& $39.91\pm0.21$ (99.77\%)
& $39.91\pm0.15$ (99.77\%)
& $78.93\pm0.80$ (98.66\%)
& $79.09\pm0.76$ (98.86\%) \\

& AG3
& $39.80\pm0.26$ (99.50\%)
& $39.50\pm0.51$ (98.75\%)
& $78.95\pm0.85$ (98.69\%)
& $78.86\pm0.97$ (98.58\%) \\
\addlinespace

6 renewable RUs (even)
& AG1
& $39.75\pm0.42$ (99.38\%)
& $39.11\pm1.05$ (97.78\%)
& $76.28\pm3.07$ (95.35\%)
& $77.30\pm2.48$ (96.63\%) \\

& AG2
& $39.49\pm0.49$ (98.73\%)
& $39.14\pm0.54$ (97.85\%)
& $79.19\pm0.76$ (98.99\%)
& $79.60\pm0.43$ (99.50\%) \\

& AG3
& $39.36\pm0.50$ (98.40\%)
& $37.57\pm1.42$ (93.92\%)
& $79.39\pm0.48$ (99.24\%)
& $74.13\pm2.81$ (92.66\%) \\
\addlinespace

8 renewable RUs
& AG1
& $39.44\pm0.58$ (98.60\%)
& $39.39\pm0.72$ (98.47\%)
& $78.11\pm1.21$ (97.64\%)
& $71.69\pm2.57$ (89.61\%) \\

& AG2
& $39.58\pm0.40$ (98.95\%)
& $39.52\pm0.37$ (98.80\%)
& $78.13\pm0.89$ (97.66\%)
& $76.73\pm1.98$ (95.91\%) \\

& AG3
& $39.71\pm0.45$ (99.28\%)
& $38.73\pm0.86$ (96.82\%)
& $79.53\pm0.41$ (99.41\%)
& $79.60\pm0.32$ (99.50\%) \\

\bottomrule
\end{tabular}
\end{table*}

Table~\ref{tab:qos_satisfaction} reports the QoS performance of TD3 and PPO under different renewable-RU deployments and traffic loads. In most scenarios, both algorithms achieve a user satisfaction rate above 95\%, indicating that the RU sleep-control policies generally maintain the required 6~Mbps user data rate. TD3 provides more stable QoS performance under the higher traffic load, with satisfaction rates ranging from 95.35\% to 99.66\%. In comparison, PPO falls below 95\% in several cases, particularly under AG3 with six evenly distributed renewable RUs and under AG1 with eight renewable RUs. These results demonstrate that the energy and carbon reductions achieved by TD3 are generally obtained while maintaining a high level of user satisfaction.

\section{Conclusion}
This paper investigated carbon-aware RU sleep control in O-RAN under heterogeneous energy supply, where renewable-powered RUs are carbon-neutral and grid-powered RUs incur operational emissions. By formulating the problem as a sequential decision-making task and embedding sustainability objectives into reward design, we systematically evaluated TD3, PPO, and heuristic control across multiple renewable penetration levels, spatial layouts, and traffic conditions.
The results reveal a clear operational transition from grid-dominant to renewable-dominant activation as renewable availability increases.
In structural terms, TD3 activates on average about one fewer RU than PPO and about 2.5 fewer RUs than the heuristic baseline, indicating a more compact and carbon-aware activation pattern. In energy terms, TD3 consistently outperforms PPO, reducing average network energy consumption by 5--10~W in the 40UE case and 5--8~W in the 80-UE case, while AG1 further achieves 2--4~W lower energy consumption than AG2 in most moderate-renewable scenarios. In carbon terms, TD3-AG2 achieves 8\%--12\% lower emissions than PPO-AG2 in low-renewable scenarios, maintains 6\%--10\% lower emissions under higher renewable penetration, and reaches up to 20\% lower emissions than the heuristic baseline under 80 UEs. In addition, AG1 reduces carbon emissions by approximately 12\%--28\% relative to AG3, with an average reduction of around 18\%, confirming that explicit carbon awareness substantially reshapes RU activation behavior.
These findings demonstrate that carbon-aware reward design is essential for sustainable O-RAN operation, and that deterministic off-policy learning provides a more effective solution for large-scale RU control under heterogeneous energy conditions.

\begin{acronym} 
\acro{5G}{Fifth Generation}
\acro{ACO}{Ant Colony Optimization}
\acro{ANN}{Artificial Neural Network}
\acro{BB}{Base Band}
\acro{BBU}{Base Band Unit}
\acro{BER}{Bit Error Rate}
\acro{BS}{Base Station}
\acro{BSs}{Base Stations}
\acro{BW}{bandwidth}
\acro{C-RAN}{Cloud Radio Access Networks}
\acro{CAPEX}{Capital Expenditure}
\acro{CoMP}{Coordinated Multipoint}
\acro{CR}{Cognitive Radio}
\acro{D2D}{Device-to-Device}
\acro{DAC}{Digital-to-Analog Converter}
\acro{DAS}{Distributed Antenna Systems}
\acro{DBA}{Dynamic Bandwidth Allocation}
\acro{DC}{Duty Cycle}
\acro{DL}{Deep Learning}
\acro{DSA}{Dynamic Spectrum Access}
\acro{FBMC}{Filterbank Multicarrier}
\acro{FEC}{Forward Error Correction}
\acro{FFR}{Fractional Frequency Reuse}
\acro{FSO}{Free Space Optics}
\acro{GA}{Genetic Algorithms}
\acro{HAP}{High Altitude Platform}
\acro{HL}{Higher Layer}
\acro{HARQ}{Hybrid-Automatic Repeat Request}
\acro{HCA}{Hierarchical Cluster Analysis}
\acro{HO}{Handover}
\acro{KNN}{k-nearest neighbors} 
\acro{IoT}{Internet of Things}
\acro{LAN}{Local Area Network}
\acro{LAP}{Low Altitude Platform}
\acro{LL}{Lower Layer}
\acro{LoS}{Line of Sight}
\acro{LTE}{Long Term Evolution}
\acro{LTE-A}{Long Term Evolution Advanced}
\acro{MAC}{Medium Access Control}
\acro{MAP}{Medium Altitude Platform}
\acro{MDP}{Markov Decision Process}
\acro{ML}{Machine Learning}
\acro{MME}{Mobility Management Entity}
\acro{mmWave}{millimeter Wave}
\acro{MIMO}{Multiple Input Multiple Output}
\acro{NFP}{Network Flying Platform}
\acro{NFPs}{Network Flying Platforms}
\acro{NLoS}{Non-Line of Sight}
\acro{OFDM}{Orthogonal Frequency Division Multiplexing}
\acro{O-RAN}{Open Radio Access Network}
\acro{OSA}{Opportunistic Spectrum Access}
\acro{PAM}{Pulse Amplitude Modulation}
\acro{PAPR}{Peak-to-Average Power Ratio}
\acro{PGW}{Packet Gateway}
\acro{PHY}{physical layer}
\acro{PSO}{Particle Swarm Optimization}
\acro{PU}{Primary User}
\acro{QAM}{Quadrature Amplitude Modulation}
\acro{QoE}{Quality of Experience}
\acro{QoS}{Quality of Service}
\acro{QPSK}{Quadrature Phase Shift Keying}
\acro{RF}{Radio Frequency}
\acro{RL}{Reinforcement Learning}
\acro{RMSE}{Root Mean Squared Error}
\acro{RN}{Remote Node}
\acro{RRH}{Remote Radio Head}
\acro{RRC}{Radio Resource Control}
\acro{RRU}{Remote Radio Unit}
\acro{RSS}{Received Signal Strength}
\acro{SU}{Secondary User}
\acro{SCBS}{Small Cell Base Station}
\acro{SDN}{Software Defined Network}
\acro{SNR}{Signal-to-Noise Ratio}
\acro{SON}{Self-organising Network}
\acro{SVM}{Support Vector Machine}
\acro{TDD}{Time Division Duplex}
\acro{TD-LTE}{Time Division LTE}
\acro{TDM}{Time Division Multiplexing}
\acro{TDMA}{Time Division Multiple Access}
\acro{UE}{User Equipment}
\acro{UAV}{Unmanned Aerial Vehicle}
\acro{USRP}{Universal Software Radio Platform}
\acro{DRL}{Deep Reinforcement Learning}
\acro{AI}{Artificial Intelligence}
\acro{RAN}{Radio Access Network}
\acro{RU}{Radio Unit}
\acro{CU}{Central Unit}
\acro{DU}{Distributed Unit}
\acro{NR}{New Radio}
\acro{gNBs}{Next Generation Node Bases}
\acro{CP}{Control Plane}
\acro{UP}{User Plane}
\acro{FPGAs}{Field  Programmable  Gate  Arrays}
\acro{ASICs}{Application-specific Integrated Circuits}
\acro{PHY-low}{lower level PHY layer processing}
\acro{FFT}{Fast Fourier Transform}
\acro{RRC}{Radio Resource Control}
\acro{SDAP}{Service Data Adaptation Protocol}
\acro{PDCP}{Packet Data Convergence Protocol}
\acro{RLC}{Radio Link Control}
\acro{RIC}{RAN Intelligent Controller}
\acro{RICs}{RAN Intelligent Controllers}
\acro{KPMs}{Key Performence Measurements}
\acro{RT}{Real Time}
\acro{SMO}{Service Management and Orchestration}
\acro{UE}{User Equipment}
\acro{API}{Application Programming Interface}
\acro{OSC}{O-RAN Software Community}
\acro{DRL}{Deep Reinforcement Learning}
\acro{OSP}{Online Service Provider}
\acro{NIB}{Network Information Base}
\acro{SDL}{Shared Data Layer}
\acro{SLA}{Service Level Agreement}
\acro{A1AP}{A1 Application Protocol}
\acro{HTTP}{Hypertext Transfer Protocol}
\acro{SL}{Supervised Learning}
\acro{UL}{Unsupervised Learning}
\acro{RL}{Reinforcement Learning}
\acro{DL}{Deep Learning}
\acro{FDD}{Frequency-division Duple}
\acro{TDD}{Time-division Duple}
\acro{LSTM}{Long Short-term Memory}
\acro{PCA}{Principal Component Analysis}
\acro{ICA}{Independent Component Analysis}
\acro{MDP}{Markov Decision Process}
\acro{GRL}{Generalization Representation Learning}
\acro{SRL}{Specialization Representation Learning}
\acro{SVM}{Support Vector Machine}
\acro{TDNN}{Time-delay Neural Network}
\acro{LSTM}{Long Short-term Memory}
\acro{MSE}{Mean Squared Error}
\acro{CNN}{Conventional Neural Network}
\acro{NAS}{Neural Architecture Search}
\acro{SDS}{Software Defined Security}
\acro{SON}{Self-organized Network}
\acro{KPIs}{Key Performance Indicators}
\acro{HetNet}{Heterogeneous Network}
\acro{HPN}{High Power Node}
\acro{LPNs}{Low Power Nodes}
\acro{QL}{Q-Learning}
\acro{PG}{Policy Gradient}
\acro{A2C}{Actor-Critic}
\acro{TD}{Temporal Defence}
\acro{SHAP}{SHapley Additive exPlanations}
\acro{DNNs}{Deep Neural Networks}
\acro{DNN}{Deep Neural Network}
\acro{MLP}{Multiple-layer Perceptron}
\acro{RNN}{Recurrent Neural Network}
\acro{DRL}{Deep Reinforcement Learning}
\acro{DQN}{Deep Q-Learning Network}
\acro{DDQN}{Double Deep Q-Learning Network}
\acro{GNN}{Graph Neural Network}
\acro{eMMB}{enhanced mobile broadband}
\acro{URLLC}{ultra-reliable low-latency communication}
\acro{QoS}{Quality of Service}
\acro{ILP}{Integer Linear Programming}
\acro{NF}{Network Function}
\acro{VFN}{Network Function Virtualization}
\acro{NBC}{Navie Bayes Classifier}
\acro{RAT}{Radio Access Technology}
\acro{FML}{federated meta-learning}
\acro{TS}{Traffic Steering}
\acro{CQL}{Conservative Q-Learning}
\acro{REM}{Random Ensemble Mixture}
\acro{SCA}{Successive Convex Approximation}
\acro{XAI}{eXplainable Artificial Intelligent}
\acro{D-RAN}{Distributed RAN}
\acro{C-RAN}{Cloud RAN}
\acro{v-RAN}{Virtual RAN}
\acro{RRH}{Remote Radio Head}
\acro{mMTC}{massive machine-type communication}
\acro{CDMA}{Code Division Multiple Access}
\acro{TDMA}{Time Division Multiple Access}
\acro{OFDMA}{Orthogonal Frequency-Division Multiple Access}
\acro{RRM}{Radio Resources Management}
\acro{NFV}{Network Function Virtualization}
\acro{D-RAN}{Distributed-RAN}
\acro{V-RAN}{Vritualized-RAN}
\acro{PA}{Power Amplifier}
\acro{SINR}{Signal to interference plus noise ratio}
\acro{mmWave}{millimeter Wave }
\acro{LOS}{Line of Sight}
\acro{NLOS}{Non-Line of Sight}
\acro{FSPL}{Free Space Path Loss}
\acro{EEMP}{Energy Efficiency Maximization Problem}
\acro{QFMEE}{QoS First Maximum EE}
\acro{SWES}{Switching-on/off based Energy Saving}
\acro{APC}{Area Power Consumption}
\acro{EE}{Energy Efficiency}
\acro{Near-RT}{Near-Real Time}
\acro{Non-RT}{Non-Real Time}
\acro{Open-RAN}{Open Radio Access Network}
\acro{Near-RT RIC}{Near Real Time RIC}
\acro{Non-RT RIC}{Non Real Time RIC}
\acro{TD3}{Twin Delayed Deep Deterministic Policy Gradient}
\acro{RC}{Radio Card}
\acro{UMa}{Urban Microcell path loss}
\acro{PRB}{Physical Resource Block}
\acro{API}{Application Programming Interface}
\acro{RSS}{Received Signal Strength}
\acro{DL}{downlink}
\acro{UL}{uplink}
\acro{DDPG}{Deep Deterministic Policy Gradient}
\acro{TD}{Temporal Difference}
\acro{RSRP}{Reference Signals Received Power}
\acro{FL}{Federated Learning}
\acro{UMi}{Urban Microcell}
\acro{BN}{Batch Normalization}
\acro{SBS}{Small Base Station}
\acro{FRL}{Federated Reinforcement Learning}
\acro{DQNSA}{DQN-Single Action}
\acro{DQNMA}{DQN-Multiple Action}
\acro{CSI}{Channel State Information}
\acro{AWGN}{Additive White Gaussian Noise}
\acro{RIS}{Reconfigurable Intelligent Surface}
\acro{PPO}{Proximal Policy Optimization}
\acro{UDN}{Ultra Dense Network}
\acro{ES}{Energy Saving}
\acro{MPC}{Model Predictive Control}
\acro{SMDP}{Semi-Markov Decision Process}
\acro{HAPS}{High Altitude Platform Station}
\end{acronym}

\bibliographystyle{IEEEtran}

\bibliography{References.bib}

@article{ju2022energy,
  title={Energy-efficient ultra-dense network with deep reinforcement learning},
  author={Ju, Hyungyu and Kim, Seungnyun and Kim, Youngjoon and Shim, Byonghyo},
  journal={IEEE Transactions on Wireless Communications},
  volume={21},
  number={8},
  pages={6539--6552},
  year={2022},
  publisher={IEEE}
}

@article{ye2019drag,
  title={DRAG: Deep reinforcement learning based base station activation in heterogeneous networks},
  author={Ye, Junhong and Zhang, Ying-Jun Angela},
  journal={IEEE Transactions on Mobile Computing},
  volume={19},
  number={9},
  pages={2076--2087},
  year={2019},
  publisher={IEEE}
}

@article{sun2024deep,
  title={Deep Learning-Based Traffic-Aware Base Station Sleep Mode and Cell Zooming Strategy in RIS-Aided Multi-Cell Networks},
  author={Sun, Shuo and Huang, Chong and Chen, Gaojie and Xiao, Pei and Tafazolli, Rahim},
  journal={IEEE Transactions on Cognitive Communications and Networking},
  year={2024},
  publisher={IEEE}
}

@misc{O-RanArchitecture,
  author = {{O-RAN Allience WG1}},
  title = {{O-RAN Architecture Description}},
  month ={June},
  year = {2025},
  note = {accessed: June 2025},
  url = {https://https://specifications.o-ran.org/specifications}
}

@misc{O-RanA1Interface,
  author = {{O-RAN Allience WG2}},
  title = {{A1 Interface: general aspects and principles 3.0}},
  month ={June},
  year = {2025},
  note = {accessed: June 2025},
  url = {https://https://specifications.o-ran.org/specifications}
}

@inproceedings{abubakar2023energy,
  title={Energy Efficiency of Open Radio Access Network: A Survey},
  author={Abubakar, Attai Ibrahim and Onireti, Oluwakayode and Sambo, Yusuf and Zhang, Lei and Ragesh, GK and Imran, Muhammad Ali},
  booktitle={2023 IEEE 97th Vehicular Technology Conference (VTC2023-Spring)},
  pages={1--7},
  year={2023},
  organization={IEEE}
}

@article{liang2024energy,
  title={Energy consumption of machine learning enhanced open RAN: A comprehensive review},
  author={Liang, Xuanyu and Wang, Qiao and Al-Tahmeesschi, Ahmed and Chetty, Swarna B and Grace, David and Ahmadi, Hamed},
  journal={IEEE Access},
  volume={12},
  pages={81889--81910},
  year={2024},
  publisher={IEEE}
}

@article{thompson2022editorial,
  title={Editorial a decade of green radio and the path to “net zero”: A United Kingdom perspective},
  author={Thompson, John S and Fletcher, Simon and Friderikos, Vasilis and Gao, Yue and Hanzo, Lajos and Nakhai, Mohammad Reza and O’Farrell, Timothy and Wells, Patricia D},
  journal={IEEE Transactions on Green Communications and Networking},
  volume={6},
  number={2},
  pages={657--664},
  year={2022},
  publisher={IEEE}
}

@article{wu2021deep,
  title={Deep reinforcement learning with spatio-temporal traffic forecasting for data-driven base station sleep control},
  author={Wu, Qiong and Chen, Xu and Zhou, Zhi and Chen, Liang and Zhang, Junshan},
  journal={IEEE/ACM transactions on networking},
  volume={29},
  number={2},
  pages={935--948},
  year={2021},
  publisher={IEEE}
}

@article{zhen2024energy,
  title={Energy-Saving Control Strategy for Ultra-Dense Network Base Stations Based on Multi-Agent Reinforcement Learning},
  author={Zhen, Yan and Tao, Litianyi and Wu, Dapeng and Tang, Tong and Wang, Ruyan},
  journal={Digital Communications and Networks},
  year={2024},
  publisher={Elsevier}
}

@article{masoudi2022digital,
  title={Digital twin assisted risk-aware sleep mode management using deep Q-networks},
  author={Masoudi, Meysam and Soroush, Ebrahim and Zander, Jens and Cavdar, Cicek},
  journal={IEEE Transactions on Vehicular Technology},
  volume={72},
  number={1},
  pages={1224--1239},
  year={2022},
  publisher={IEEE}
}

@article{gan2025joint,
  title={Joint Sleep Control and Energy Sharing Strategy With Deep Reinforcement Learning in Green Ultra-Dense Networks},
  author={Gan, Jipeng and Kou, Hongxia and Yang, Guang and Zhang, Hua and Cao, Zijian and Xu, Wei},
  journal={IEEE Transactions on Green Communications and Networking},
  year={2025},
  publisher={IEEE}
}

@techreport{3gppTR38901,
  author       = {{3GPP}},
  title        = {Study on channel model for frequencies from 0.5 to 100 GHz},
  institution  = {ETSI},
  number       = {TR 138 901},
  type         = {Technical Report},
  version      = {14.3.0},
  year         = {2017},
  url          = {https://www.etsi.org/deliver/etsi_tr/138900_138999/138901.pdf}
}

@misc{O-RanE2Interface,
  author = {{O-RAN Allience WG3}},
  title = {{O-RAN E2 General Aspects and Principles (E2GAP) 8.0}},
  month ={October},
  year = {2025},
  note = {accessed: October 2025},
  url = {https://https://specifications.o-ran.org/specifications}
}

@article{lin2021data,
  title={A data-driven base station sleeping strategy based on traffic prediction},
  author={Lin, Jiansheng and Chen, Youjia and Zheng, Haifeng and Ding, Ming and Cheng, Peng and Hanzo, Lajos},
  journal={IEEE Transactions on Network Science and Engineering},
  volume={11},
  number={6},
  pages={5627--5643},
  year={2021},
  publisher={IEEE}
}

@article{zhou2023joint,
  title={Joint user association and bs operation for green communications in ultra-dense heterogeneous networks},
  author={Zhou, Tianqing and Fu, Yanyan and Qin, Dong and Li, Xuan and Li, Chunguo},
  journal={IEEE Transactions on Vehicular Technology},
  volume={73},
  number={2},
  pages={2305--2319},
  year={2023},
  publisher={IEEE}
}

@article{gong2014base,
  title={Base station sleeping and resource allocation in renewable energy powered cellular networks},
  author={Gong, Jie and Thompson, John S and Zhou, Sheng and Niu, Zhisheng},
  journal={IEEE Transactions on Communications},
  volume={62},
  number={11},
  pages={3801--3813},
  year={2014},
  publisher={IEEE}
}

@article{wu2013traffic,
  title={Traffic-aware base station sleeping control and power matching for energy-delay tradeoffs in green cellular networks},
  author={Wu, Jian and Zhou, Sheng and Niu, Zhisheng},
  journal={IEEE Transactions on Wireless Communications},
  volume={12},
  number={8},
  pages={4196--4209},
  year={2013},
  publisher={IEEE}
}

@article{sun2025flexible,
  title={Flexible Base Station Sleeping and Resource Allocation for Green Uplink Fully-Decoupled RAN},
  author={Sun, Yu and Zhou, Haibo and Yu, Kai and Xu, Yunting and Qian, Bo and Cai, Lin X},
  journal={IEEE Transactions on Wireless Communications},
  year={2025},
  publisher={IEEE}
}

@article{wu2021smdp,
  title={SMDP-based sleep policy for base stations in heterogeneous cellular networks},
  author={Wu, Jing and Li, Yun and Zhuang, Hongcheng and Pan, Zhiwen and Wang, Guoyin and Xian, Yongju},
  journal={Digital Communications and Networks},
  volume={7},
  number={1},
  pages={120--130},
  year={2021},
  publisher={Elsevier}
}

@article{lee2019adaptive,
  title={Adaptive traffic management and energy cooperation in renewable-energy-powered cellular networks},
  author={Lee, Hyun-Suk and Lee, Jang-Won},
  journal={IEEE Systems Journal},
  volume={14},
  number={1},
  pages={132--143},
  year={2019},
  publisher={IEEE}
}

@article{baldini2024toward,
  title={Toward sustainable O-RAN deployment: An in-depth analysis of power consumption},
  author={Baldini, Gianmarco and Bolla, Raffaele and Bruschi, Roberto and Carrega, Alessandro and Davoli, Franco and Lombardo, Chiara and Rabbani, Ramin},
  journal={IEEE Transactions on Green Communications and Networking},
  year={2024},
  publisher={IEEE}
}

@article{salamatmoghadasi2023energy,
  title={Energy sustainability in dense radio access networks via high altitude platform stations},
  author={Salamatmoghadasi, Maryam and Mehrabian, Amir and Yanikomeroglus, Halim},
  journal={IEEE Networking Letters},
  volume={6},
  number={1},
  pages={21--25},
  year={2023},
  publisher={IEEE}
}

@inproceedings{renga2025network,
  title={Network Sharing: a pathway to Sustainability and Carbon Footprint Mitigation in Radio Access Networks},
  author={Renga, Daniela and Meo, Michela and Nuaymi, Loutfi},
  booktitle={2025 36th International Teletraffic Congress (ITC-36)},
  pages={1--9},
  year={2025},
  organization={IEEE}
}

@article{ahmadi2025towards,
  title={Towards Sustainability in 6G and beyond: Challenges and Opportunities of Open RAN},
  author={Ahmadi, Hamed and Rahmani, Mostafa and Chetty, Swarna Bindu and Tsiropoulou, Eirini Eleni and Arslan, Huseyin and Debbah, Merouane and Quek, Tony},
  journal={IEEE Communications Standards Magazine},
  year={2025},
  publisher={IEEE}
}

@article{liang2026scalable,
  author={Liang, Xuanyu and Al-Tahmeesschi, Ahmed and Chetty, Swarna and Cavdar, Cicek and Canberk, Berk and Ahmadi, Hamed},
  journal={IEEE Transactions on Green Communications and Networking}, 
  title={Scalable Machine Learning-Based Approaches for Energy Saving in Densely Deployed Open RAN}, 
  year={2026},
  volume={10},
  pages={2710-2722},
  doi={10.1109/TGCN.2026.3679951}}

@article{DING2022106339,
title = {Carbon emissions and mitigation potentials of 5G base station in China},
journal = {Resources, Conservation and Recycling},
volume = {182},
pages = {106339},
year = {2022},
issn = {0921-3449},
doi = {https://doi.org/10.1016/j.resconrec.2022.106339},
url = {https://www.sciencedirect.com/science/article/pii/S092134492200177X},
author = {Yangxue Ding and Huabo Duan and Minghui Xie and Ruichang Mao and JiaJia Wang and Wenbo Zhang},
}
\end{document}